\documentclass[aps,preprint,amsmath,amssymb,superscriptaddress,showpacs,nofootinbib]{revtex4}
\usepackage{multirow}
\usepackage{graphicx}
\usepackage{float}
\usepackage{caption}
\usepackage{subcaption}
\usepackage{slashed}
\begin{document}

\title{Investigation of anomalous triple gauge couplings in $\mu\gamma$ collision at multi-TeV muon colliders}

\author{S. Spor}
\email[]{serdar.spor@beun.edu.tr}
\affiliation{Department of Medical Imaging Techniques, Zonguldak B\"{u}lent Ecevit University, 67100, Zonguldak, Turkey.}

\author{M. K{\"o}ksal}
\email[]{mkoksal@cumhuriyet.edu.tr} 
\affiliation{Department of Physics, Sivas Cumhuriyet University, 58140, Sivas, Turkey.}

\begin{abstract}
The pursuit of discovery in particle physics has always required study at the highest possible energies. Multi-TeV muon colliders, which have important advantages, offer unprecedented potential for investigating new physics beyond the Standard Model. We extensively assume a range of center-of-mass energies from 3 to 30 TeV and a range of integrated luminosities from 1 to 90 ab$^{-1}$ for future multi-TeV muon colliders. We perform a phenomenological study of the anomalous $WW\gamma$ couplings via the process $\mu^+ \mu^-\,\rightarrow\,\mu^+\gamma^* \mu^-\,\rightarrow\,\mu^+\ell^-\nu_\ell \bar{\nu}_\ell$ using a model-independent analysis in the effective field theory framework at multi-TeV muon colliders. The sensitivity estimates obtained for the anomalous $WW\gamma$ couplings at systematic uncertainties of $\delta_{sys}=0$, $3$, $5\%$ are compared with the experimental results. Our best sensitivity limits at 95$\%$ confidence level for $c_{WWW}/{\Lambda^2}$, $c_{\widetilde{W}WW}/\Lambda^2$, $c_{B}/{\Lambda^2}$, $c_{W}/{\Lambda^2}$ and $c_{\widetilde{W}}/\Lambda^2$ couplings are $[-0.030; 0.025]\,\text{TeV}^{-2}$, $[-0.014; 0.014]\,\text{TeV}^{-2}$, $[-0.185; 0.187]\,\text{TeV}^{-2}$, $[-0.187; 0.189]\,\text{TeV}^{-2}$ and $[-1.177; 1.097]\,\text{TeV}^{-2}$ at the multi-TeV muon collider with center-of-mass energy of 30 TeV and integrated luminosity of 90 ab$^{-1}$, respectively.
\end{abstract}

\pacs{12.60.-i, 14.70.Hp, 14.70.Bh \\
Keywords: Electroweak Interaction, Models Beyond the Standard Model, Anomalous Triple Gauge Couplings, Muon Collider.\\}

\maketitle

\section{Introduction}

The Standard Model (SM) of elementary particle physics, validated by the highest-energy and most sensitive experiments available to date, provides the best description of the universe. However, some open questions need to be answered, such as the nature of dark matter, neutrino oscillations, matter-antimatter asymmetry of the universe, and the strong $CP$ problem. Therefore, the necessity of searching for new physics beyond the SM has emerged. The absence of any clues at the Large Hadron Collider (LHC), although so far it has great potential for discovering new physics, makes it clear that a new multi-TeV collider may be needed to definitively test what lies beyond the SM.

A new multi-TeV muon collider has several key advantages over lepton and hadron colliders. Since the proton is a composite particle and the muon is a fundamental particle, muon-muon collisions are cleaner than proton-proton collisions. At the same time, the mass of the muon is about 207 times heavier than that of the electron. Due to the greater mass of the muon, muon beams produce less synchrotron radiation than electron beams. They can be easily accelerated at high energies in this manner. To mention another important advantage, electron-positron collisions require the use of a linear collider and a single detector to produce collisions with significant luminosity at very high energies, while muon collisions can be accelerated in multi-pass rings and detected with multiple detectors \cite{Boscolo:2019ytr}. The ability to use multi-pass rings in muon colliders is more efficient and more compact than electron-positron linear colliders, resulting in significant power and cost savings. As a result, the future muon collider is a significant candidate to be an ideal discovery machine to complement the LHC by providing precise measurements with clean channels at multi-TeV energy ranges above the LHC. The motivation to build a future muon collider, which obviously can answer many of the fundamental questions in particle physics, is increasing day by day with the Muon Accelerator Program (MAP) \cite{Palmer:2014asd,Delahaye:2013hvz}, Muon Ionization Cooling Experiment (MICE) \cite{Bogomilov:2020twm} and Low Emittance Muon Accelerator (LEMMA) \cite{Antonelli:2016ezx} studies. Recently, many possible phenomenological studies beyond the SM have been offered for multi-TeV muon colliders in \cite{Buttazzo:2018wmg,Costantini:2020tkp,Chiesa:2020yhn,Bandyopadhyay:2021lja,Han:2021hrq,Liu:2021gtr,Han:2021twq,Capdevilla:2021xku,Bottaro:2021res,Capdevilla:2021ooc,Yin:2020gre,Ruhdorfer:2020tgx,Huang:2021edc,Asadi:2021wsd,Han:2021pas}.

The effective Lagrangian method in a model-independent manner based on Effective Field Theory (EFT) is a convenient and popular approach to defining possible new physics effects. In this article, we have studied the anomalous triple gauge coupling (aTGC) beyond the SM using the effective Lagrangian method with $SU(2)\times U(1)$ invariant dimension-6 effective operators \cite{Sahin:2011dfg,Kumar:2015ghe,Etesami:2016eto,Bian:2016wer,Ari:2016aac,Falkowski:2017ghe,Sahin:2017uot,Li:2018tyb,Bhatia:2019gso,Rahaman:2020abs}. We focus on the anomalous $WW\gamma$ coupling in the photon-induced process at the multi-TeV muon collider. A few years ago, $W^+W^-$, $t\bar{t}$ and $t\bar{t}H$ productions in two-photon induced processes at 30 TeV muon collider have been investigated using the effective photon approximation \cite{Han:2021opx}, we detail this approximation in Section 2. Therefore, to specify the most suitable muon collider, we consider various center-of-mass energies from 3 to 30 TeV and integrated luminosities increasing in proportion to \cite{Franceschini:2021pol}

\begin{eqnarray}
\label{eq.1} 
{\cal L}_{\text{int}}=10\,\text{ab}^{-1}\left(\frac{\sqrt{s}}{10\,\text{TeV}}\right)^2.
\end{eqnarray}

Corresponding to the center-of-mass energies, the energies of incoming muon and the integrated luminosities are given in Table~\ref{tab1} and these values for the muon collider have been used in previous studies \cite{Han:2021twq,Ali:2021wsd,Chiesa:2021tyr}.

\begin{table}[H]
\centering
\caption{The used values of muon collider.}
\label{tab1}
\begin{tabular}{p{3cm}p{1.5cm}p{1.5cm}p{1.5cm}p{1.5cm}p{1.5cm}}
\hline \hline
$\sqrt{s}$ (TeV) & 3 & 6 & 10 & 14 & 30\\ \hline
$E_{\mu}$ (TeV) & 1.5 & 3 & 5 & 7 & 15\\ \hline
${\cal L}_{\text{int}}$ (ab$^{-1}$) & 1 & 4 & 10 & 20 & 90\\  \hline \hline
\end{tabular}
\end{table}

The effective Lagrangian allows us to investigate possible deviations from the predictions of the SM, assuming $SU(2)\times U(1)$ symmetry of the electroweak gauge field. The SM Lagrangian ${\cal L}_{\text{SM}}$ consists only of dimension-4 operators, and higher dimensional operators express the new physics contributions beyond the SM. These operators are suppressed by the inverse powers of the new physics scale ${\Lambda}$. Therefore, the larger the dimension of the operator, the smaller the new physics effects. The effective Lagrangian can be written as

\begin{eqnarray}
\label{eq.2} 
{\cal L}_{\text{eft}}={\cal L}_{\text{SM}}+\sum_{i}\frac{c_i^{(6)}}{\Lambda^{2}}{\cal O}_i^{(6)}+\text{h.c.}\,,
\end{eqnarray}

{\raggedright where $c_i^{(6)}$ is the coupling of dimension-6 operator ${\cal O}_i^{(6)}$. The effective Lagrangian with the dimension-6 operators describing the low-energy effects of the new physics is \cite{Hagiwara:1993aso,Degrande:2013rry}}

\begin{eqnarray}
\label{eq.3} 
{\cal L}_{\text{eft}}=\frac{1}{\Lambda^2}\left[C_W{\cal O}_W+C_B{\cal O}_B+C_{WWW}{\cal O}_{WWW}+C_{\widetilde{W}WW}{\cal O}_{\widetilde{W}WW}+C_{\widetilde{W}}{\cal O}_{\widetilde{W}}+h.c.\right]\,,
\end{eqnarray}

{\raggedright where dimension-6 operators}

\begin{eqnarray}
\label{eq.4} 
{\cal O}_{WWW}=\text{Tr}\left[W_{\mu\nu}W^{\nu\rho}W_\rho^\mu\right]\,,
\end{eqnarray}
\begin{eqnarray}
\label{eq.5} 
{\cal O}_{W}=\left(D_\mu\Phi\right)^\dagger W^{\mu\nu}\left(D_\nu\Phi\right)\,,
\end{eqnarray}
\begin{eqnarray}
\label{eq.6} 
{\cal O}_{B}=\left(D_\mu\Phi\right)^\dagger B^{\mu\nu}\left(D_\nu\Phi\right)\,,
\end{eqnarray}
\begin{eqnarray}
\label{eq.7} 
{\cal O}_{\widetilde{W}WW}=\text{Tr}\left[ \widetilde{W}_{\mu\nu}W^{\nu\rho}W_\rho^\mu\right]\,,
\end{eqnarray}
\begin{eqnarray}
\label{eq.8} 
{\cal O}_{\widetilde{W}}=\left(D_\mu\Phi\right)^\dagger \widetilde{W}^{\mu\nu}\left(D_\nu\Phi\right)\,.
\end{eqnarray}

Here, $\Phi$ is the Higgs doublet field. ${\cal O}_{WWW}$, ${\cal O}_{W}$ and ${\cal O}_{B}$ are three $CP$-conserving operators while ${\cal O}_{\widetilde{W}WW}$ and ${\cal O}_{\widetilde{W}}$ are two $CP$-violating operators. $D_\mu$ is the covariant derivative and $B^{\mu\nu}$ and $W^{\mu\nu}$ are the field strength tensors of $SU\left(2\right)$ and $U\left(1\right)$ gauge fields defined as

\begin{eqnarray}
\label{eq.9} 
D_\mu \equiv \partial_\mu\,+\,i\frac{g^\prime}{2}B_\mu\,+\,igW_\mu^i\frac{\tau^i}{2}\,,
\end{eqnarray}
\begin{eqnarray}
\label{eq.10} 
B_{\mu\nu}=\frac{i}{2}g^\prime\left(\partial_\mu B_\nu - \partial_\nu B_\mu\right)\,,
\end{eqnarray}
\begin{eqnarray}
\label{eq.11} 
W_{\mu\nu}=\frac{i}{2}g\tau^i\left(\partial_\mu W_\nu^i - \partial_\nu W_\mu^i + g\epsilon_{ijk}W_\mu^j W_\nu^k\right)\,.
\end{eqnarray}

The anomalous $WW\gamma$ coupling is conventionally parametrized with terms of the effective Lagrangian given by \cite{Hagiwara:1987xdj,Bian:2015ylk}:

\begin{eqnarray}
\label{eq.12} 
{\cal L}_{\text{eft}}^{WW\gamma}&=&ig_{WW\gamma}\Big[{g_1^{\gamma}}\left({W_{\mu\nu}^{+}}{W_{\mu}^{-}}A_{\nu}-{W_{\mu\nu}^{-}}{W_{\mu}^{+}}A_{\nu}\right) \nonumber \\
&+&\kappa_{\gamma}{W_{\mu}^{+}}{W_{\nu}^{-}}A_{\mu\nu}+\frac{\lambda_{\gamma}}{M_W^2}{W_{\mu\nu}^{+}}{W_{\nu\rho}^{-}}A_{\rho\mu} \nonumber \\
&+&ig_4^\gamma {W_{\mu}^{+}}{W_{\nu}^{-}}\left( \partial_\mu A_\nu+\partial_\nu A_\mu\right) \\
&-&ig_5^\gamma \epsilon_{\mu\nu\rho\sigma}\left({W_{\mu}^{+}}\partial_\rho{W_{\nu}^{-}}-\partial_\rho{W_{\mu}^{+}}{W_{\nu}^{-}}\right)A_\sigma \nonumber \\
&+&\widetilde{\kappa}_{\gamma}{W_{\mu}^{+}}{W_{\nu}^{-}}\tilde{A}_{\mu\nu}+\frac{\widetilde{\lambda}_{\gamma}}{M_W^2}{W_{\lambda\mu}^{+}}{W_{\mu\nu}^{-}}\tilde{A}_{\nu\lambda} \Big]\,, \nonumber 
\end{eqnarray}

{\raggedright where $A^{\mu\nu}=\partial^\mu A^\nu - \partial^\nu A^\mu$, $W_{\mu\nu}^\pm=\partial_\mu W_\nu^\pm - \partial_\nu W_\mu^\pm$, $\tilde{A}_{\mu\nu}=\frac{1}{2}\epsilon_{\mu\nu\rho\sigma}A^{\rho\sigma}$ and the overall coupling contant is given by $g_{WW\gamma}=-e$. The Lagrangian in Eq.~(\ref{eq.12}) has 7 free $WW\gamma$ couplings. Due to electromagnetic gauge invariance, three of these couplings are ${g_1^{\gamma}}=1$ and $g_4^\gamma=g_5^\gamma=0$. However, $\kappa_{\gamma}$ and $\lambda_{\gamma}$ conserve $CP$ while $\widetilde{\kappa}_{\gamma}$ and $\widetilde{\lambda}_{\gamma}$ violate $CP$. In the SM, $\kappa_{\gamma}=1$ ($\Delta \kappa_{\gamma}=0$). Here, this coupling is at tree level and 1-loop
corrections have been calculated in Refs.~\cite{Couture:1987ecp,Couture:1988yvw}. Finally, all other couplings are zero. The couplings of the Lagrangian in Eq.~(\ref{eq.12}) can be derived in terms of the couplings of the operators in Eqs.~(\ref{eq.4})-(\ref{eq.8}) as \cite{Hagiwara:1993aso,Wudka:1994abs,Degrande:2013rry}}

\begin{eqnarray}
\label{eq.13} 
{\kappa_\gamma}=1+\Delta \kappa_{\gamma}=1+\left(c_W+c_B\right)\frac{m_W^2}{2\Lambda^2}\,,
\end{eqnarray}
\begin{eqnarray}
\label{eq.14} 
{\lambda_\gamma}=c_{WWW}\frac{3g^2m_W^2}{2\Lambda^2}\,,
\end{eqnarray}
\begin{eqnarray}
\label{eq.15} 
\widetilde{\kappa}_{\gamma}=c_{\widetilde{W}}\frac{m_W^2}{2\Lambda^2}\,,
\end{eqnarray}
\begin{eqnarray}
\label{eq.16} 
\widetilde{\lambda}_{\gamma}=c_{\widetilde{W}WW}\frac{3g^2m_W^2}{2\Lambda^2}\,.
\end{eqnarray}

In this study, we focus on the anomalous $c_{WWW}/{\Lambda^2}$, $c_{W}/{\Lambda^2}$ and $c_{B}/{\Lambda^2}$ coupling parameters that are $CP$-conserving and $c_{\widetilde{W}WW}/\Lambda^2$ and $c_{\widetilde{W}}/\Lambda^2$ coupling parameters that are $CP$-violating. In the SM, the anomalous coupling parameters are given by $c_{WWW}/\Lambda^2=c_{W}/\Lambda^2=c_{B}/\Lambda^2=0$. Any change in the values of the anomalous $c_{WWW}/\Lambda^2$, $c_{W}/\Lambda^2$, $c_{B}/\Lambda^2$ parameters creates a deviation from the SM corresponding to the new physics contributions for the $WW\gamma$ couplings \cite{Spor:2020tgn,Spor:2021omb,Spor:2022elk,Koksal:2020gbm,Rodriguez:2020ukz,Billur:2021yqa}.

The current most sensitive limits on the anomalous $WW\gamma$ couplings have been derived by CMS and ATLAS Collaborations at the LHC with center-of-mass energy of 13 TeV. The limits of the anomalous coupling parameters on the aTGC obtained in the latest experimental studies are given in Table~\ref{tab2}.

\begin{table}
\caption{The experimental observed limits at 95$\%$ C.L. on the aTGC with $c_{WWW}/\Lambda^2$, $c_{\widetilde{W}WW}/{\Lambda^2}$, $c_{B}/\Lambda^2$, $c_{W}/\Lambda^2$ and $c_{\widetilde{W}}/{\Lambda^2}$ parameters in recent years.}
\label{tab2}
\begin{ruledtabular}
\begin{tabular}{lccccc}
\multirow{2}{*}{Experimental limit} & $c_{WWW}/\Lambda^2$ & $c_{\widetilde{W}WW}/\Lambda^2$ & $c_{B}/\Lambda^2$ & $c_{W}/\Lambda^2$ & $c_{\widetilde{W}}/\Lambda^2$\\ 
& (TeV$^{-2}$) & (TeV$^{-2}$) & (TeV$^{-2}$) & (TeV$^{-2}$) & (TeV$^{-2}$)\\
\hline
CMS Collaboration  \cite{Sirunyan:2021rwx} & [-0.90; 0.91] & [-0.45; 0.45] & [-40.0; 41.0] & -- & [-20.0; 20.0]\\
CMS Collaboration  \cite{Sirunyan:2019umc} & [-1.58; 1.59] & -- & [-8.78; 8.54] & [-2.00; 2.65] & --\\
ATLAS Collaboration \cite{Aaboud:2019abc} & [-3.40; 3.30] & [-1.60; 1.60] & [-21.0; 18.0] & [-7.40; 4.10] & [-76.0; 76.0]\\
ATLAS Collaboration \cite{Aaboud:2017les} & [-3.10; 3.10] & -- & [-19.0; 20.0] & [-5.10; 5.80] & --\\
\end{tabular}
\end{ruledtabular}
\end{table}

\section{Cross-sections at the multi-TeV muon colliders}

\begin{figure}[H]
\centering
\includegraphics{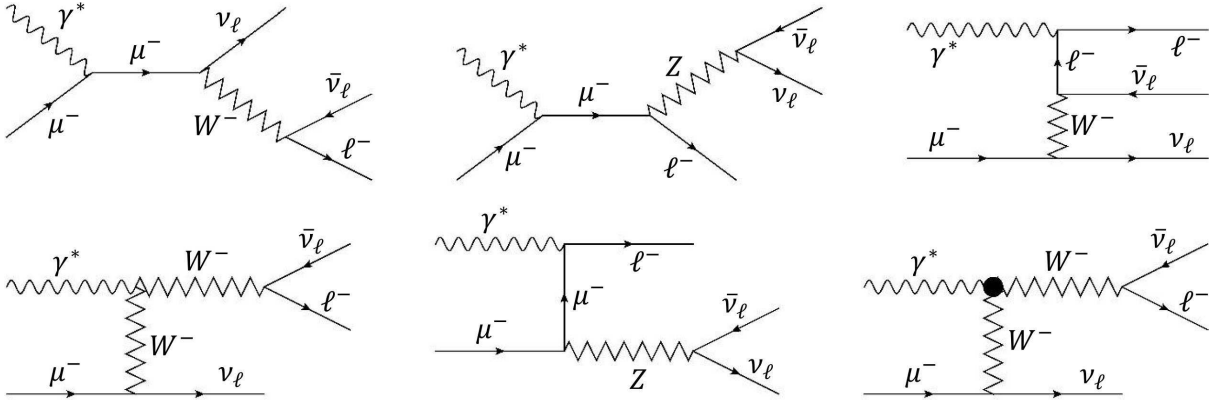}
\caption{The Feynman diagrams of the subprocess $\mu^-\gamma^*\,\rightarrow\,\ell^-\nu_\ell \bar{\nu}_\ell$. 
\label{fig1}}
\end{figure}

The $\mu^-\gamma^*\,\rightarrow\,\ell^-\nu_\ell \bar{\nu}_\ell$ participates as a subprocess in the main process $\mu^+ \mu^-\,\rightarrow\,\mu^+\gamma^* \mu^-\,\rightarrow\,\mu^+\ell^-\nu_\ell \bar{\nu}_\ell$ ($\ell^-=e^-,\mu^-$). The Feynman diagrams for the subprocess $\mu^-\gamma^*\,\rightarrow\,\ell^-\nu_\ell \bar{\nu}_\ell$ are given in Fig.~\ref{fig1}. Tau is the most heavily charged lepton and has a lifetime of $3\times10^{-13}$ s by decaying into lighter leptons, electron, muon and lighter hadrons such as $\pi$'s and $K$'s. Primary decay channels can be presented by one charged particle (one prong decay) and three charged particle (three prong decay). Of all the tau decays, 85$\%$ are the one prong decays and 15$\%$ are the three prong decays. Particles produced from tau decays are called tau jets. One prong lepton jets are directly identified by similar algorithms used by electron and muon, while identification of hadronic jets is more complicated than leptonic modes due to the QCD jets as background. Also, tau jets are highly collimated and are distinguished from background because of its topology \cite{Atag:2010xcv}. It's easy to study decays to electrons and muons together but reconstructing decays to taus needs a completely different approach. For this reason, we assume that the $\ell^-=e^-,\,\mu^-$.

The emitted quasi-real photon $\gamma^*$ in the $\mu^-\gamma^*$ interaction is scattered with small angles from the beam pipe of $\mu^+$ and has a low virtuality \cite{Koksal:2019ybm}. These photons are described by the Equivalent Photon Approximation (EPA) \cite{Budnev:1975kyp,Baur:2002hjh,Piotrzkowski:2001tvz}.

In this study, the photon $\gamma^*$ emitted from the muon interacts with another muon. However, processes such as $\mu^-\gamma^*\,\rightarrow\,\ell^-\nu_\ell \bar{\nu}_\ell$ are called the photon-induced processes. The spectrum of photon emitted from muon is given by \cite{Budnev:1975kyp,Koksal:2019ybm}:

\begin{eqnarray}
\label{eq.17}
\begin{aligned}
f_{\gamma^{*}}(x)=&\, \frac{\alpha}{\pi E_{\mu}}\Bigg\{\left[\frac{1-x+x^{2}/2}{x}\right]\text{log}\left(\frac{Q_{\text{max}}^{2}}{Q_{\text{min}}^{2}}\right)-\frac{m_{\mu}^{2}x}{Q_{\text{min}}^{2}}\left(1-\frac{Q_{\text{min}}^{2}}{Q_{\text{max}}^{2}}\right)
\\
&-\frac{1}{x}\left[1-\frac{x}{2}\right]^{2}\text{log}\left(\frac{x^{2}E_{\mu}^{2}+Q_{\text{max}}^{2}}{x^{2}E_{\mu}^{2}+Q_{\text{min}}^{2}}\right)\Bigg\}\,,
\end{aligned}
\end{eqnarray}

{\raggedright where $x=E_{\gamma^{*}}/E_{\mu}$. $Q_{\text{max}}^{2}$ is maximum virtuality of the photon. The minimum value of $Q_{\text{min}}^{2}$ is given as} 

\begin{eqnarray}
\label{eq.18}
Q_{\text{min}}^{2}=\frac{m_{\mu}^{2}x^{2}}{1-x}\,.
\end{eqnarray}

The total cross section of the process $\mu^+ \mu^-\,\rightarrow\,\mu^+\gamma^* \mu^-\,\rightarrow\,\mu^+\ell^-\nu_\ell \bar{\nu}_\ell$ is obtained by integrating over both the cross section of the subprocess $\mu^-\gamma^*\,\rightarrow\,\ell^-\nu_\ell \bar{\nu}_\ell$ and the spectrum of photon emitted from muon. The total cross section can be written as follows:

\begin{eqnarray}
\label{eq.19}
\sigma\left( \mu^+ \mu^-\,\rightarrow\,\mu^+\gamma^* \mu^-\,\rightarrow\,\mu^+\ell^-\nu_\ell \bar{\nu}_\ell \right)=\int f_{\gamma^{*}}(x)\hat{\sigma}\left({\mu^-\gamma^*\,\rightarrow\,\ell^-\nu_\ell \bar{\nu}_\ell}\right) dx\,.
\end{eqnarray}

In this study, the signal and the background events analysis of the photon-induced process at muon collider are simulated using {\sc MadGraph5}$\_$aMC@NLO \cite{Alwall:2014cvc}. We have enabled this program to run via the EWdim6 model file for the operators that examine interactions between the electroweak gauge boson described above in dimension-6. 

The short lifetime of the muons at rest causes some experimental difficulty. The muons will decay all within the machine, and the muon decay products will inevitably interact with its environment, creating a large number of secondary and tertiary particles that can reach the detector. However, such produced particles lead to ``beam-induced background'' (BIB) which negatively affects the potential of multi-TeV muon colliders \cite{Benedetto:2018uyt,Bartosik:2020ewn}. Although reducing the detector's coverage, the backgrounds can be removed by placing the shielding nozzles on the detector. It is considered that the effect of BIB can be neglected if the detector has coverage only up to the pseudorapidity $|\eta|\leq2.5$ and the transverse momentum is $p_T\geq25$ GeV \cite{Asadi:2021wsd}. Therefore, the kinematic cuts have been applied to achieve high signal efficiency by suppressing the backgrounds. 

In this study, the variations of pseudorapidity of the charged leptons $|\eta^\ell|$, missing energy transverse $\slashed{E}_T$ and the transverse momentums of the charged leptons $p_T^\ell$ kinematic cuts are discussed. We apply $|\eta^\ell| < 2.5$ and $p_T^\ell >10$ GeV with tagged Cut-1, also these cuts are minimum cuts at the generator level. With the same cuts of $|\eta^\ell|$ and $p_T^\ell$ in Cut-1, only $\slashed{E}_T > 50$ GeV with tagged Cut-2 is added and then by just changing values of $p_T^\ell$ cuts according to Cut-2, we also apply $p_T^\ell > 200$ GeV with tagged Cut-3. In Table~\ref{tab3}, we list applied kinematic cuts for the analysis.

\begin{table}
\centering
\caption{List of applied kinematic cuts for the analysis.}
\label{tab3}
\begin{tabular}{p{3cm}p{10cm}}
\hline \hline
Cuts & Definitions \\ 
\hline
Cut-1 & $|\eta^\ell| < 2.5$ and $p_T^\ell > 10$ GeV\\ 
Cut-2 & Same as in Cut-1, in addition $\slashed{E}_T > 50$ GeV\\ 
Cut-3 & Same as in Cut-2, but for $p_T^\ell > 200$ GeV\\ \hline \hline
\end{tabular}
\end{table}

The calculated cross sections from applied cuts can be presented in Table~\ref{tab4} to determine the efficiency of each cut in the analysis of signals and backgrounds. Anomalous couplings for signal processes are set in five different ways; $c_{WWW}/{\Lambda^2}=3$ TeV$^{-2}$ and the other four being zero, $c_{\widetilde{W}WW}/\Lambda^2=3$ TeV$^{-2}$ and the other four being zero, $c_{B}/{\Lambda^2}=40$ TeV$^{-2}$ and the other four being zero, $c_{W}/{\Lambda^2}=40$ TeV$^{-2}$ and the other four being zero, $c_{\widetilde{W}}/{\Lambda^2}=40$ TeV$^{-2}$ and the other four being zero. Some cuts required in the background processes are used as follows: transverse momentum $p_T^{\gamma}>10$ GeV, pseudo-rapidity $|\eta^{\gamma}|<2.5$, minimum distance between leptons $\Delta R^{\ell\ell}_{\text{min}}>0.4$ and minimum distance between photons and leptons $\Delta R^{\gamma \ell}_{\text{min}}>0.4$. 

The coordinate system in the detector is the right-handed Cartesian coordinate system with its origin at the interaction point (IP) in the center of the detector. The $x-$axis points from the IP to the center of collider ring, the $y-$axis points upward and the $z-$axis points to the beam direction. Cylindrical coordinates are used in the transverse plane where $\phi$ is the azimuthal angle around the beam pipe, and the $\theta$ is the polar angle from the beam axis. The pseudorapidity is identified in terms of the polar angle $\theta$ as $\eta=-\text{ln}\,\text{tan}(\theta/2)$. The $\Delta R_{\ell\ell}$ is the distance between the two charged leptons in $(\eta, \phi)$ space defined as $\Delta R_{\ell\ell}=\sqrt{\Delta\eta^2_{\ell\ell}+\Delta\phi^2_{\ell\ell}}$ where $\Delta\eta_{\ell\ell}$ and $\Delta\phi_{\ell\ell}$ represent the difference in pseudorapidity and in azimuthal angle between the two charged leptons, respectively.

The final effect of two different cuts applied after Cut-1 at the generator level are about $11.42\%$ for $c_{WWW}/{\Lambda^2}$, $26.13\%$ for $c_{\widetilde{W}WW}/\Lambda^2$, $5.44\%$ for $c_{B}/\Lambda^2$, $5.42\%$ for $c_{W}/\Lambda^2$ and $6.86\%$ for $c_{\widetilde{W}}/\Lambda^2$. On the other hand, efficiency of cuts for backgrounds ($B1-B4$) are $4.76\%$, $49.48\%$, $8.32\%$ and $61.55\%$, respectively. As a result, these cuts, which are determined to reduce the backgrounds and increase the observability of the signals, seem to be effective.

\begin{table}
\caption{Cross sections for the signals and the relevant backgrounds according to the applied cuts for the muon collider with center-of-mass energy of 14 TeV.}
\label{tab4}
\begin{ruledtabular}
\begin{tabular}{lccc}
& & Cross sections (pb) & \\ \hline
Signals & Cut-1 & Cut-2 & Cut-3\\ \hline
$c_{WWW}/{\Lambda^2}=3$ TeV$^{-2}$ & 2.635 & 1.362 & 0.301\\ 
$c_{\widetilde{W}WW}/\Lambda^2=3$ TeV$^{-2}$ & 3.276 & 1.949 & 0.856\\ 
$c_{B}/{\Lambda^2}=40$ TeV$^{-2}$ & 2.171 & 1.034 & 0.118\\ 
$c_{W}/{\Lambda^2}=40$ TeV$^{-2}$ & 2.176 & 1.041 & 0.118\\ 
$c_{\widetilde{W}}/\Lambda^2=40$ TeV$^{-2}$ & 2.595 & 1.290 & 0.178\\
\hline
Backgrounds & Cut-1 & Cut-2 & Cut-3\\  \hline
B1 ($\ell^-\nu_\ell \bar{\nu}_\ell$) & 2.438 & 1.160 & 0.116\\ 
B2 ($\ell^-\ell^-\ell^+$) & 6.279$\times 10^{-4}$ & no $\slashed{E}_T$ & 3.107$\times 10^{-4}$ without $\slashed{E}_T$\\ 
B3 ($\ell^-\nu_\ell \bar{\nu}_\ell \gamma$) & 7.033$\times 10^{-2}$ & 4.725$\times 10^{-2}$ & 5.853$\times 10^{-3}$\\  
B4 ($\ell^-\ell^-\ell^+ \gamma$) & 3.714$\times 10^{-5}$ & no $\slashed{E}_T$ & 2.286$\times 10^{-5}$ without $\slashed{E}_T$\\ 
\end{tabular}
\end{ruledtabular}
\end{table}

\begin{figure}
\centering
\begin{subfigure}{0.5\linewidth}
\includegraphics[scale=0.5]{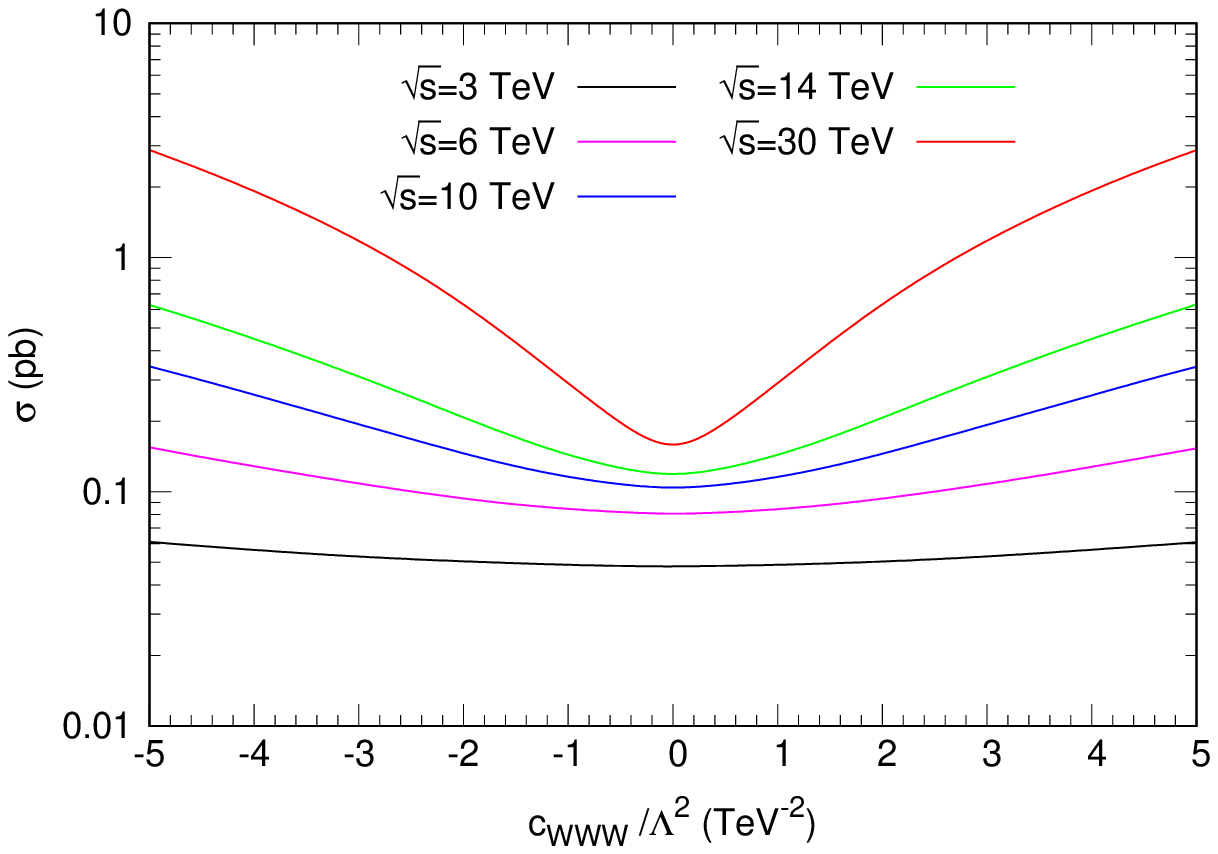}
\caption{}
\label{fig2:a}
\end{subfigure}\hfill
\begin{subfigure}{0.5\linewidth}
\includegraphics[scale=0.5]{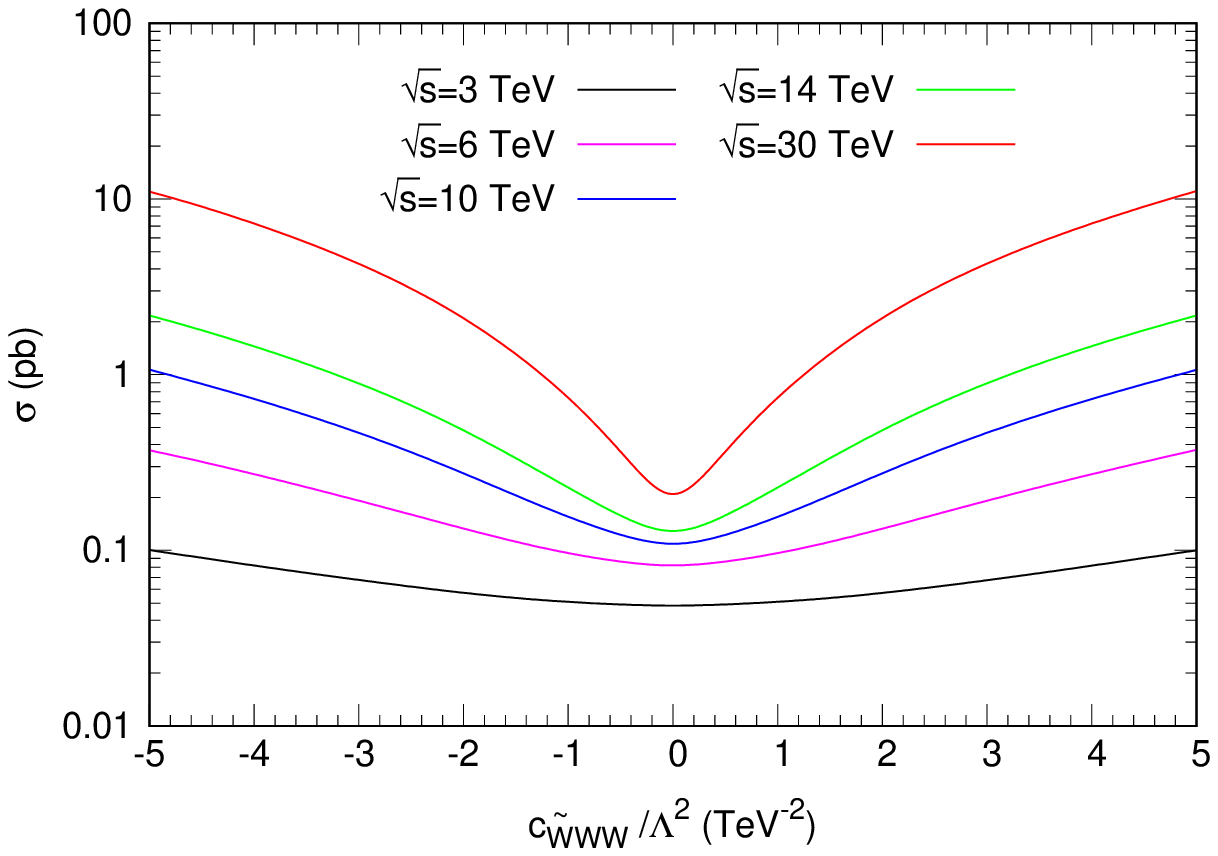}
\caption{}
\label{fig2:b}
\end{subfigure}\hfill

\begin{subfigure}{0.5\linewidth}
\includegraphics[scale=0.5]{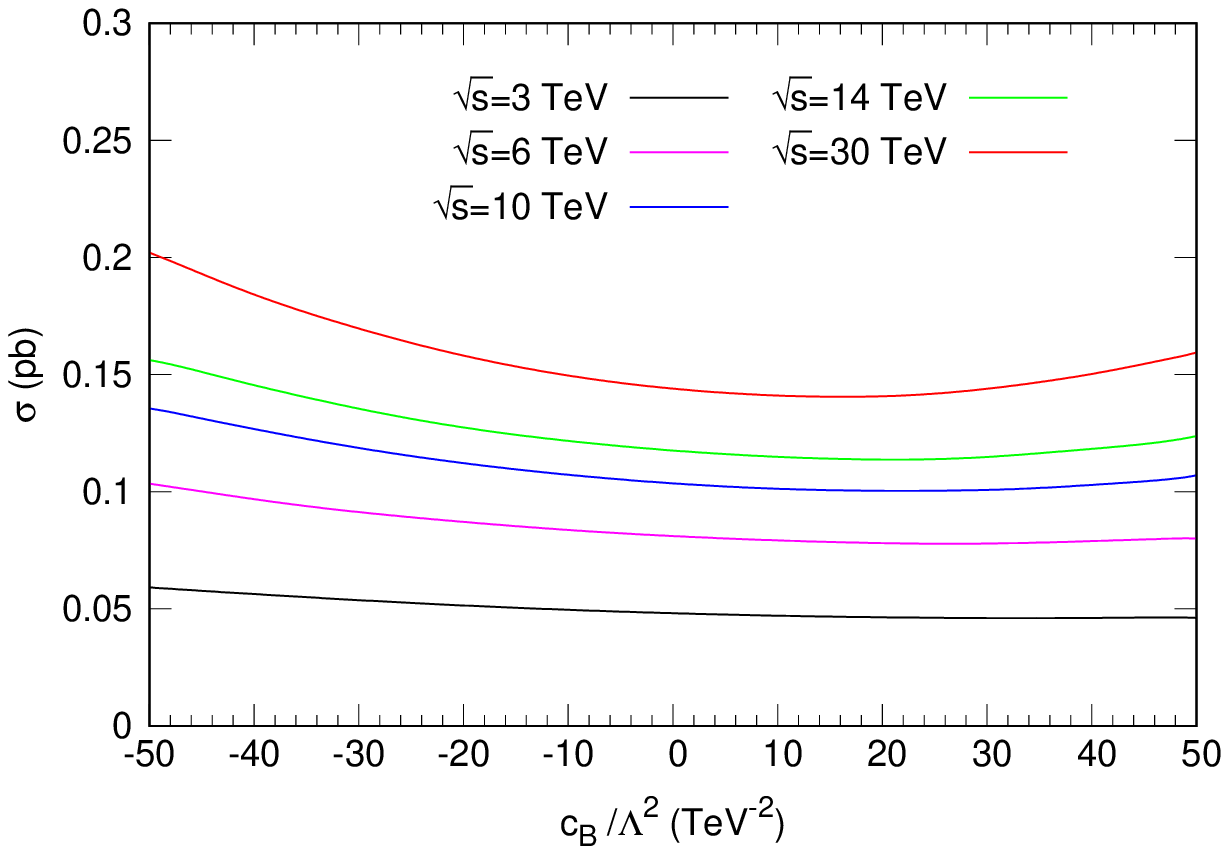}
\caption{}
\label{fig2:c}
\end{subfigure}\hfill
\begin{subfigure}{0.5\linewidth}
\includegraphics[scale=0.5]{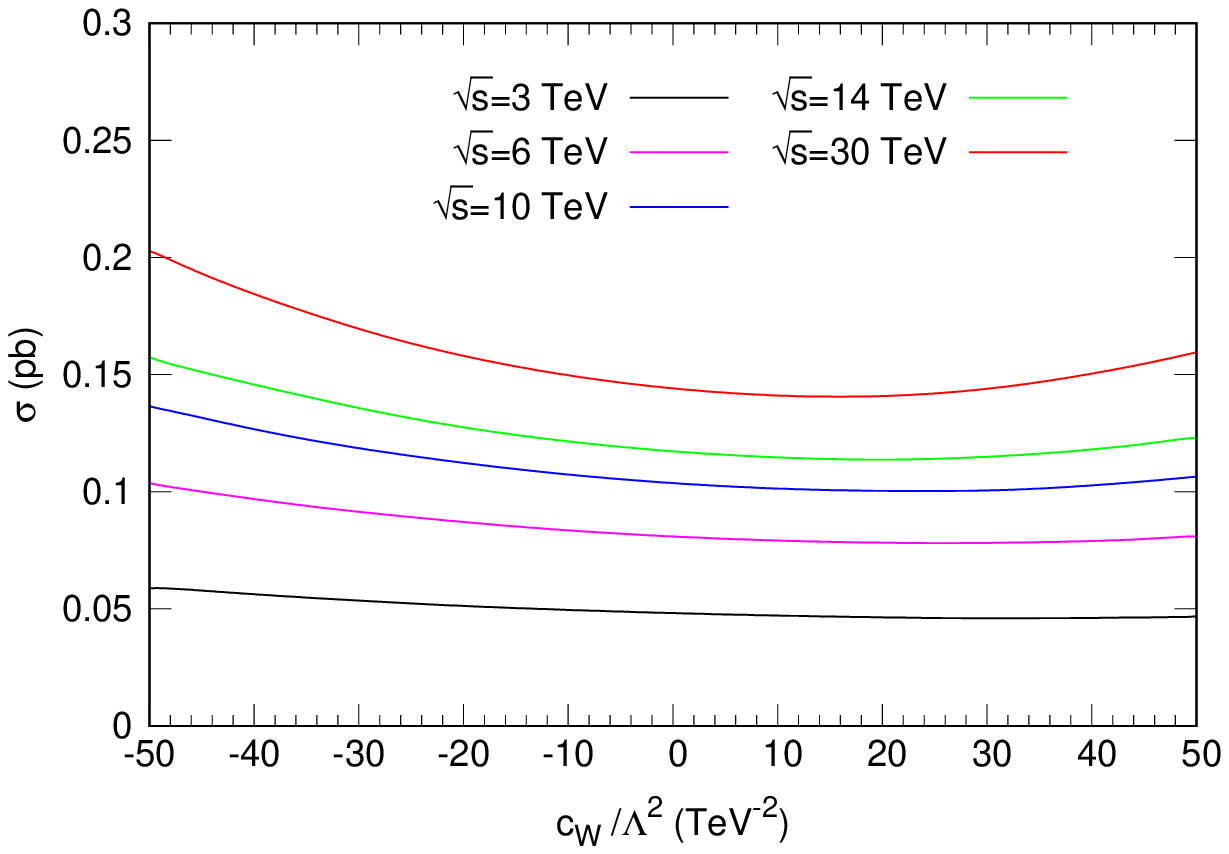}
\caption{}
\label{fig2:d}
\end{subfigure}\hfill

\begin{subfigure}{0.5\linewidth}
\includegraphics[scale=0.5]{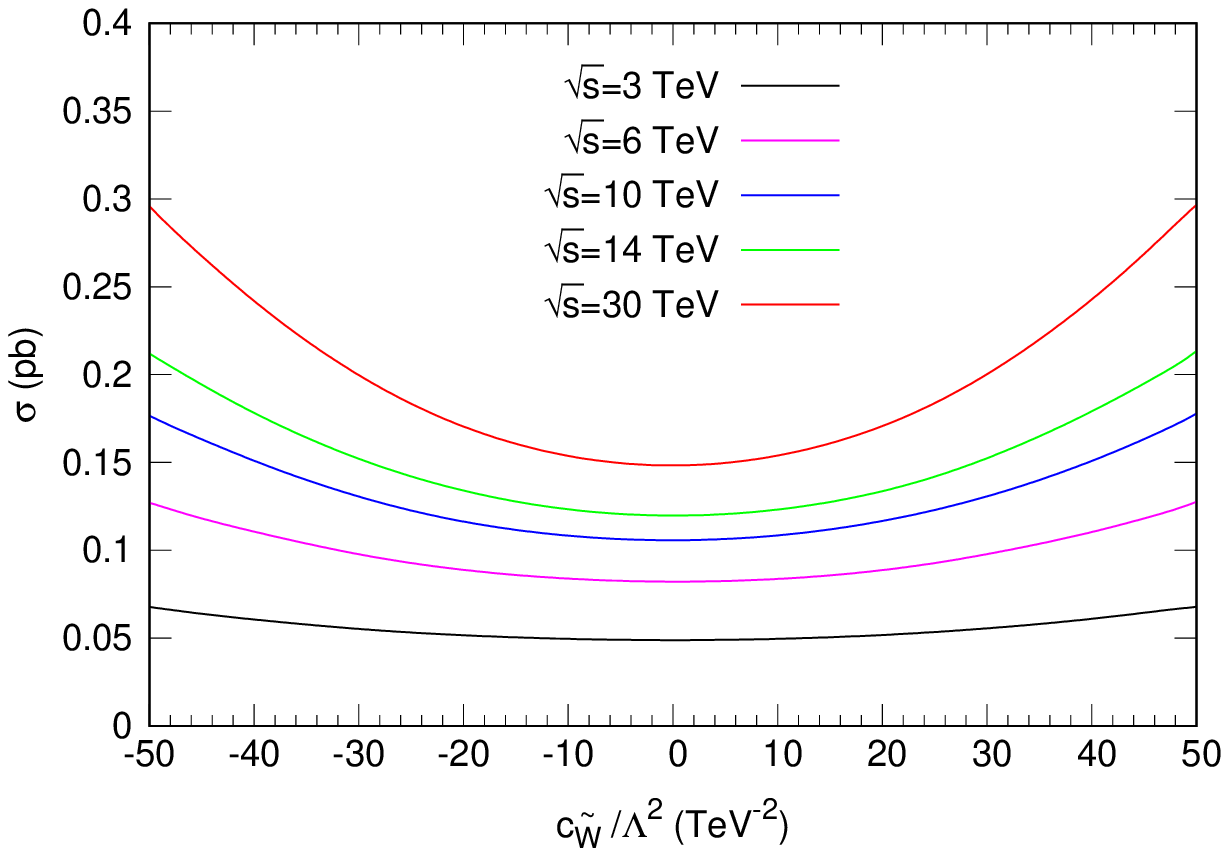}
\caption{}
\label{fig2:e}
\end{subfigure}\hfill
\caption{The total cross sections of the main process $\mu^+ \mu^-\,\rightarrow\,\mu^+\gamma^* \mu^-\,\rightarrow\,\mu^+\ell^-\nu_\ell \bar{\nu}_\ell$ as a function of the anomalous (a) $c_{WWW}/{\Lambda^2}$, (b) $c_{\widetilde{W}WW}/\Lambda^2$, (c) $c_{B}/{\Lambda^2}$, (d) $c_{W}/{\Lambda^2}$ and (e) $c_{\widetilde{W}}/\Lambda^2$ couplings for center-of-mass energies of $\sqrt{s}=3$ TeV, 6 TeV, 10 TeV, 14 TeV and 30 TeV.}
\label{fig2}
\end{figure}

In Fig.~\ref{fig2}, the total cross sections in the main process $\mu^+ \mu^-\,\rightarrow\,\mu^+\gamma^* \mu^-\,\rightarrow\,\mu^+\ell^-\nu_\ell \bar{\nu}_\ell$ as a function of the anomalous $c_{WWW}/{\Lambda^2}$, $c_{\widetilde{W}WW}/\Lambda^2$, $c_{B}/{\Lambda^2}$, $c_{W}/{\Lambda^2}$ and $c_{\widetilde{W}}/\Lambda^2$ couplings are presented for Cut-3 at the muon colliders with $\sqrt{s}=3$, $6$, $10$, $14$ and $30$ TeV. Because of Eq.~(\ref{eq.13}), the total cross section curves of the $c_{W}/\Lambda^2$ and $c_{B}/\Lambda^2$ couplings in Figs.~\ref{fig2:c}-\ref{fig2:d} show similar properties and have values close to each other. In the analyzes made for all center-of-mass energies of the muon collider, when the cross sections of the anomalous couplings with the same coupling value and center-of-mass energy are examined, the cross sections for $c_{\widetilde{W}WW}/\Lambda^2$ coupling are higher than those of other couplings. In addition, as the center-of-mass energies of the muon collider increase from 3 to 30 TeV in Fig.~\ref{fig2}, the cross sections for each anomalous coupling also increase. If we focus on the relationship between new physics contribution and center-of-mass energies, the increase in the center-of-mass energy in all analyzes also increases the difference between the SM cross section and the total cross section. Thus, the new physics effect can be seen more clearly in multi-TeV muon colliders.

\begin{figure}
\centering
\begin{subfigure}{0.5\linewidth}
\includegraphics[scale=0.6]{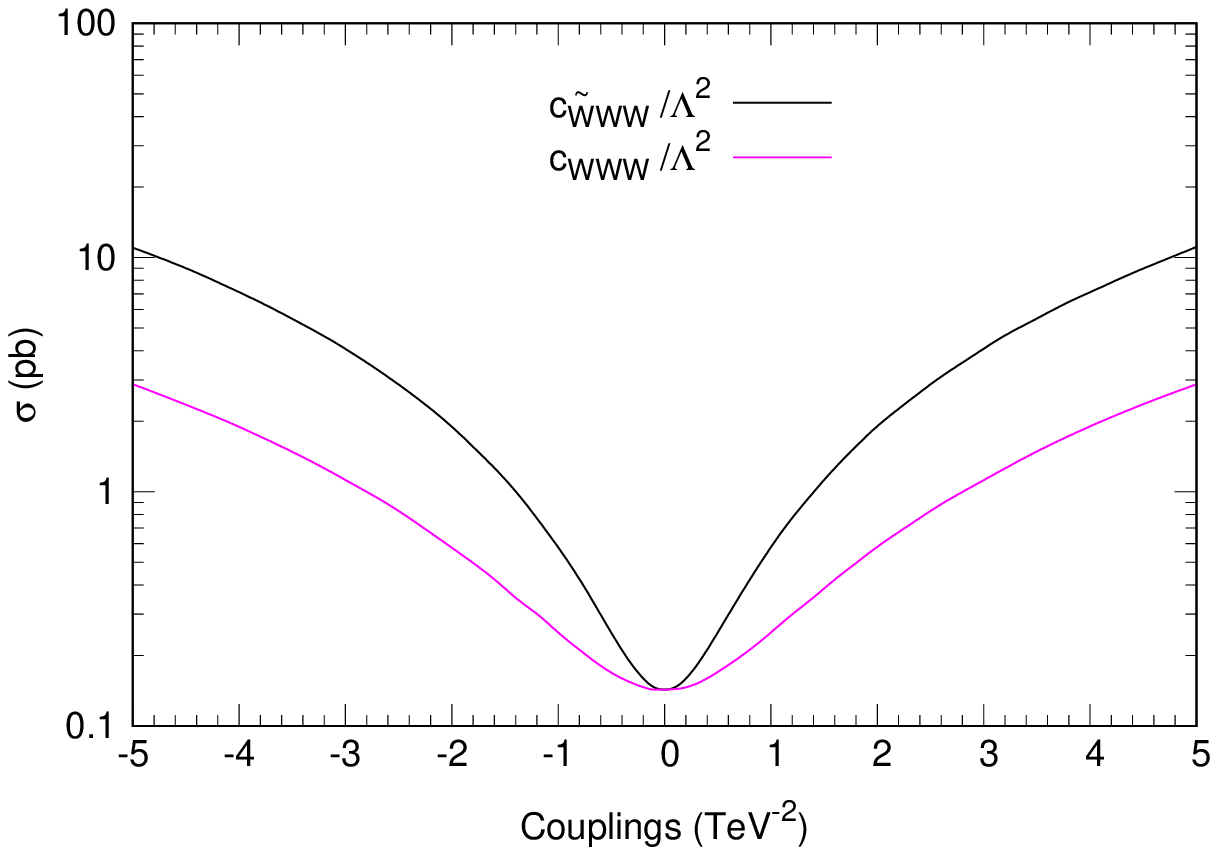}
\caption{}
\label{fig3:a}
\end{subfigure}\hfill
\begin{subfigure}{0.5\linewidth}
\includegraphics[scale=0.6]{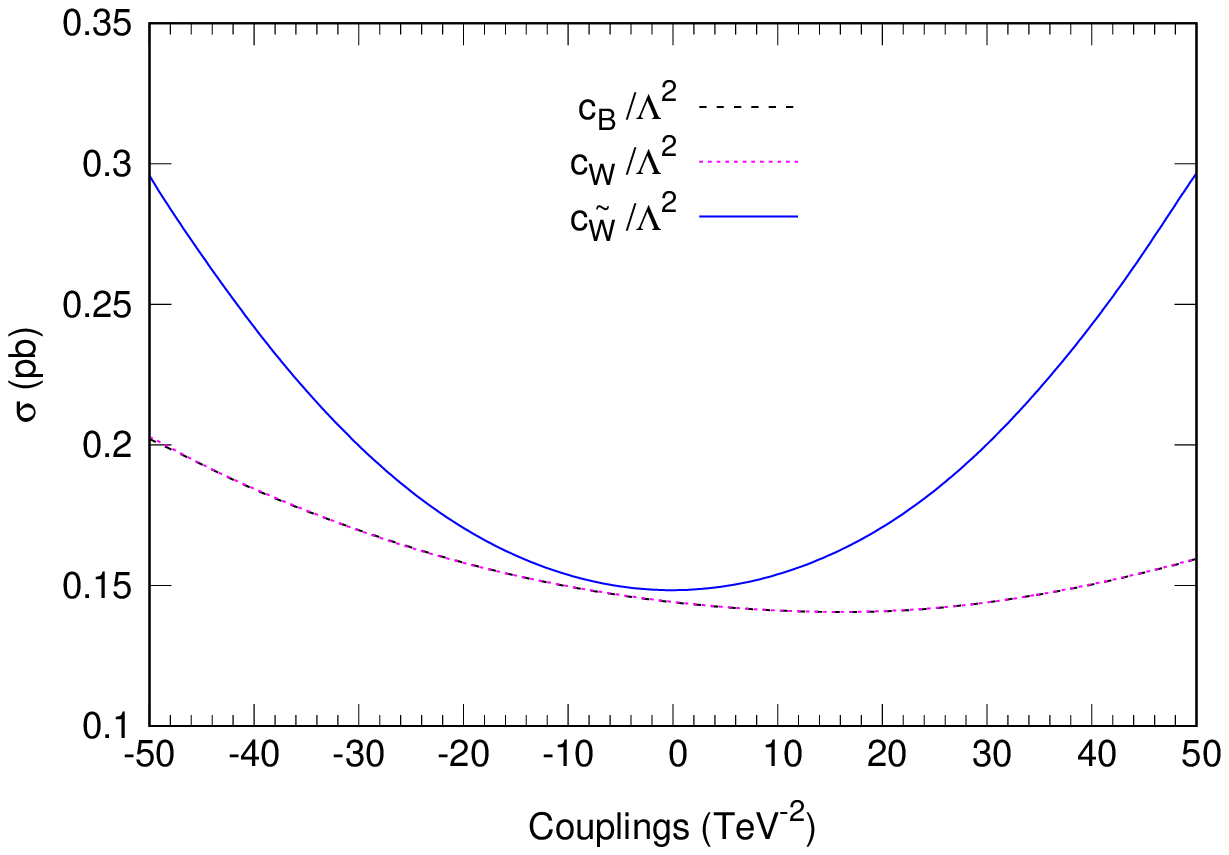}
\caption{}
\label{fig3:b}
\end{subfigure}\hfill
\caption{Comparing the total cross sections of the main process $\mu^+ \mu^-\,\rightarrow\,\mu^+\gamma^* \mu^-\,\rightarrow\,\mu^+\ell^-\nu_\ell \bar{\nu}_\ell$ as a function of the anomalous (a) $c_{WWW}/{\Lambda^2}$ and $c_{\widetilde{W}WW}/\Lambda^2$, (b) $c_{B}/{\Lambda^2}$, $c_{W}/{\Lambda^2}$ and $c_{\widetilde{W}}/\Lambda^2$ couplings at $\sqrt{s}=30$ TeV.}
\label{fig3}
\end{figure}

Fig.~\ref{fig3} aims to compare the anomalous couplings with each other at center-of-mass energy of $\sqrt{s}=30$ TeV, where the highest total cross-sections are obtained. In Fig.~\ref{fig3:a}, the signal rate of anomalous $c_{\widetilde{W}WW}/\Lambda^2$ coupling is higher than that of anomalous $c_{WWW}/{\Lambda^2}$ coupling, and in Fig.~\ref{fig3:b}, the signal rate of anomalous $c_{\widetilde{W}}/\Lambda^2$ coupling is higher than that of anomalous $c_{B}/{\Lambda^2}$ and $c_{W}/{\Lambda^2}$ coupling. This result reveals that CP-violating and CP-conserving interactions can be sharply distinguished from each other.

\section{Signal and background analysis}

The process $\mu^+ \mu^-\,\rightarrow\,\mu^+\gamma^* \mu^-\,\rightarrow\,\mu^+\ell^-\nu_\ell \bar{\nu}_\ell$ is considered as signal including SM contribution as well as interference between the anomalous $c_{WWW}/{\Lambda^2}$, $c_{\widetilde{W}WW}/\Lambda^2$, $c_{B}/{\Lambda^2}$, $c_{W}/{\Lambda^2}$ and $c_{\widetilde{W}}/\Lambda^2$ couplings and SM contributions ($S + B1$). Since the final state of the signal is ${\ell^-}\nu_{\ell}\bar{\nu}_{\ell}$, all relevant backgrounds are selected in accordance with this final state. As relevant backgrounds, we consider not only the SM contribution ($B1$) with the same final state of the signal process, but also $\ell^-\ell^-\ell^+$, ${\ell^-}\nu_\ell \bar{\nu}_\ell \gamma$ and $\ell^-\ell^-\ell^+ \gamma$. We have labelled these backgrounds as: $B2$ ($\mu^-\gamma^* \,\rightarrow\,\ell^-\ell^-\ell^+$), $B3$ ($\mu^-\gamma^* \,\rightarrow\,{\ell^-}\nu_\ell \bar{\nu}_\ell \gamma$) and $B4$ ($\mu^-\gamma^* \,\rightarrow\,\ell^-\ell^-\ell^+ \gamma$) processes.

\begin{figure}
\centering
\begin{subfigure}{0.5\linewidth}
\includegraphics[width=\linewidth]{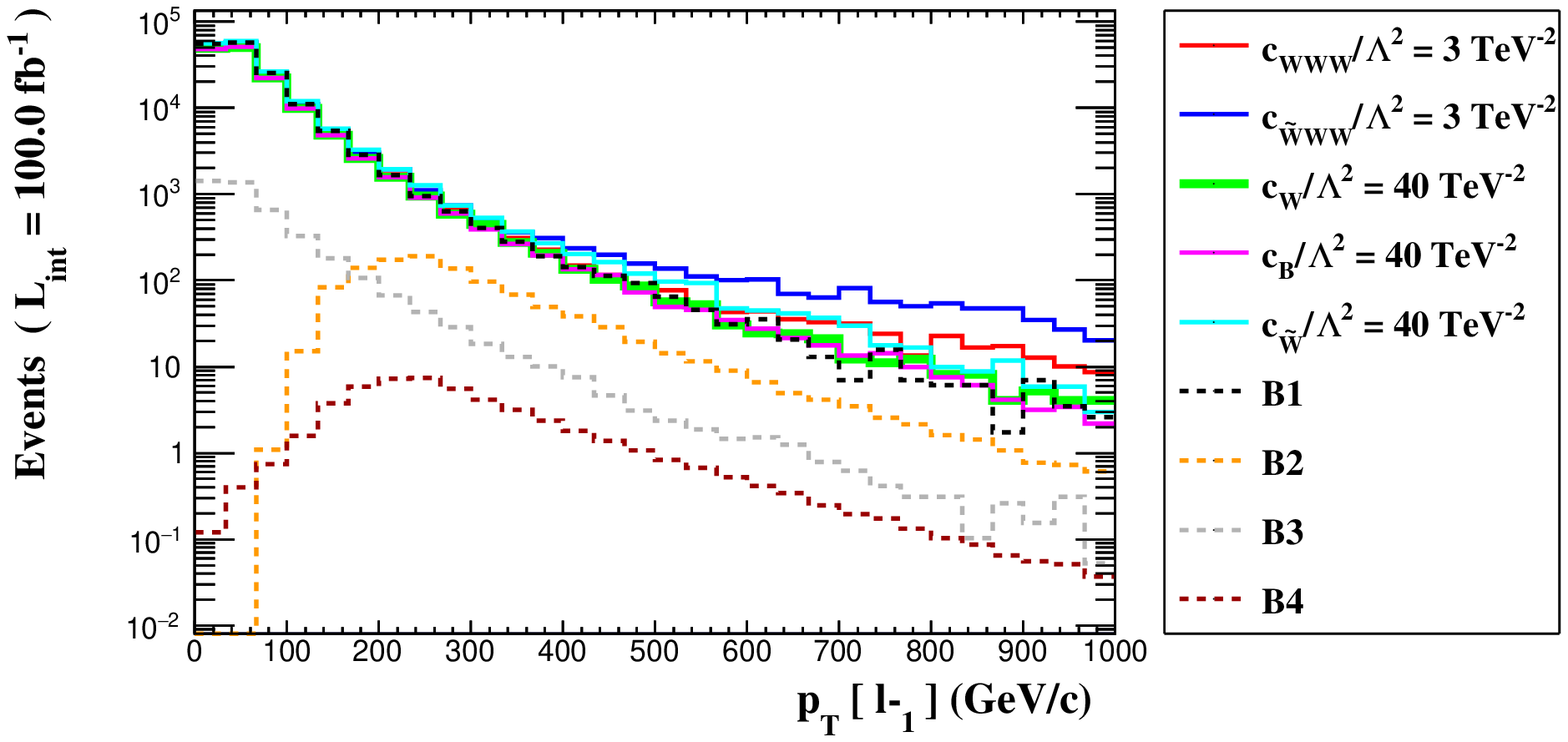}
\caption{}
\label{fig4:a}
\end{subfigure}\hfill
\begin{subfigure}{0.5\linewidth}
\includegraphics[width=\linewidth]{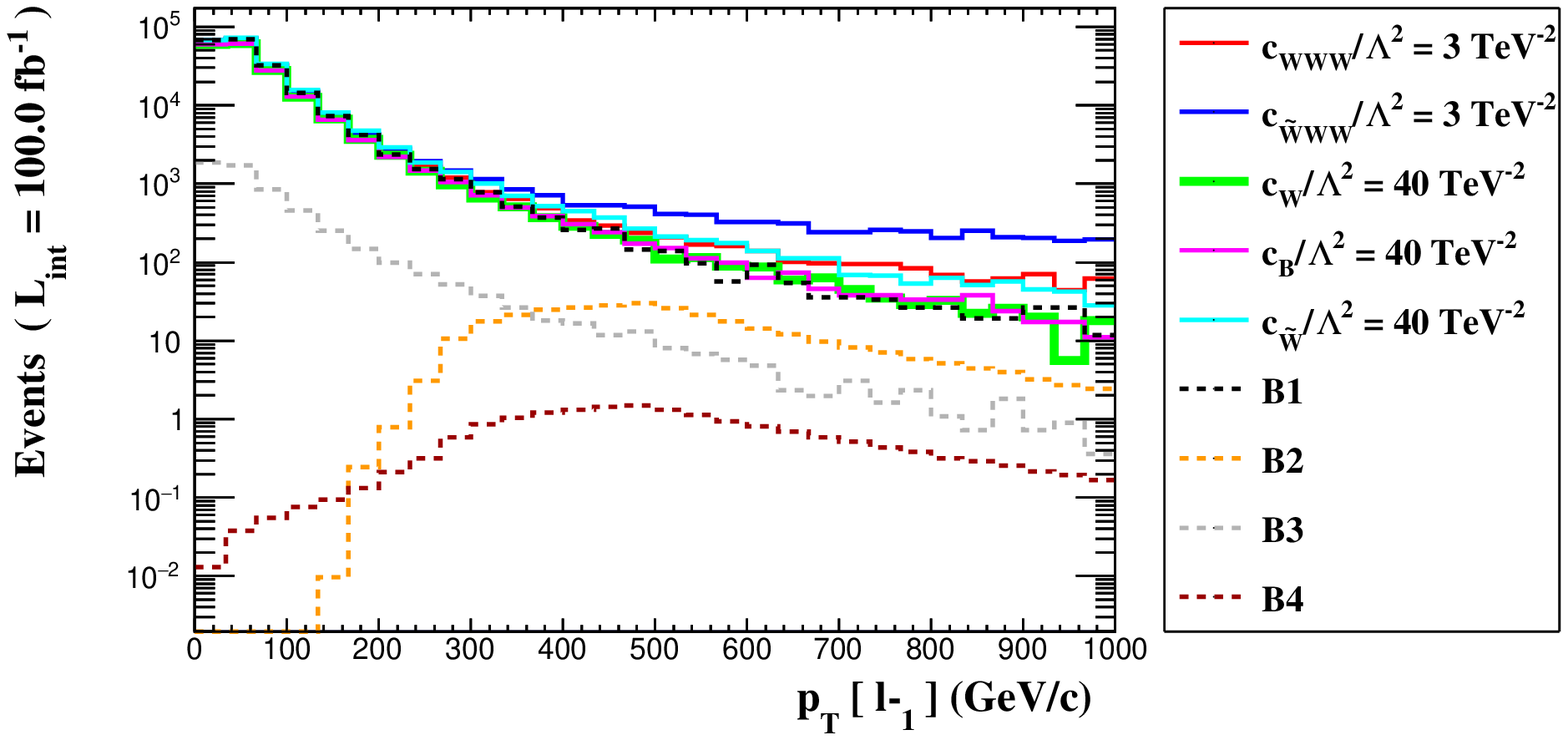}
\caption{}
\label{fig4:b}
\end{subfigure}\hfill

\begin{subfigure}{0.5\linewidth}
\includegraphics[width=\linewidth]{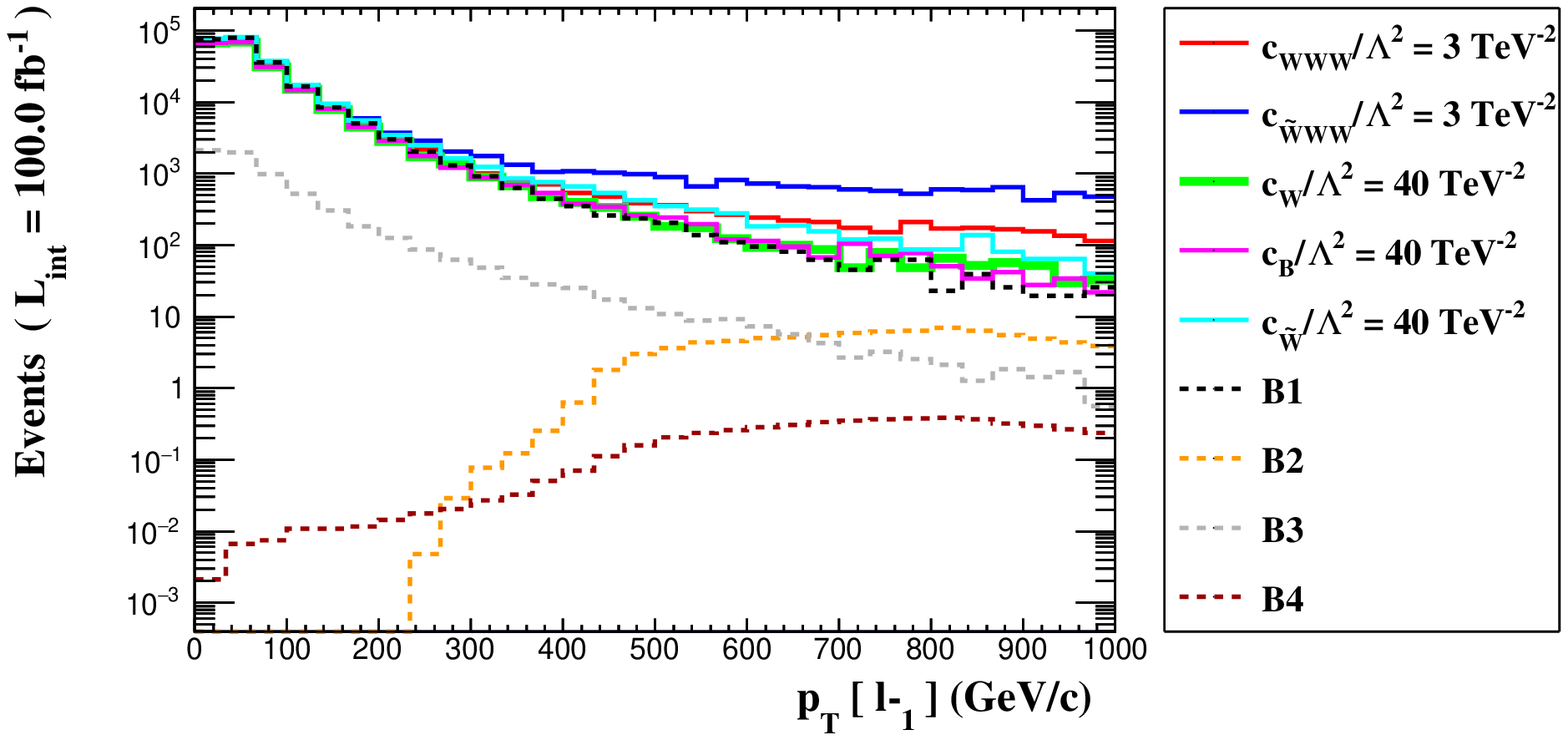}
\caption{}
\label{fig4:c}
\end{subfigure}\hfill
\begin{subfigure}{0.5\linewidth}
\includegraphics[width=\linewidth]{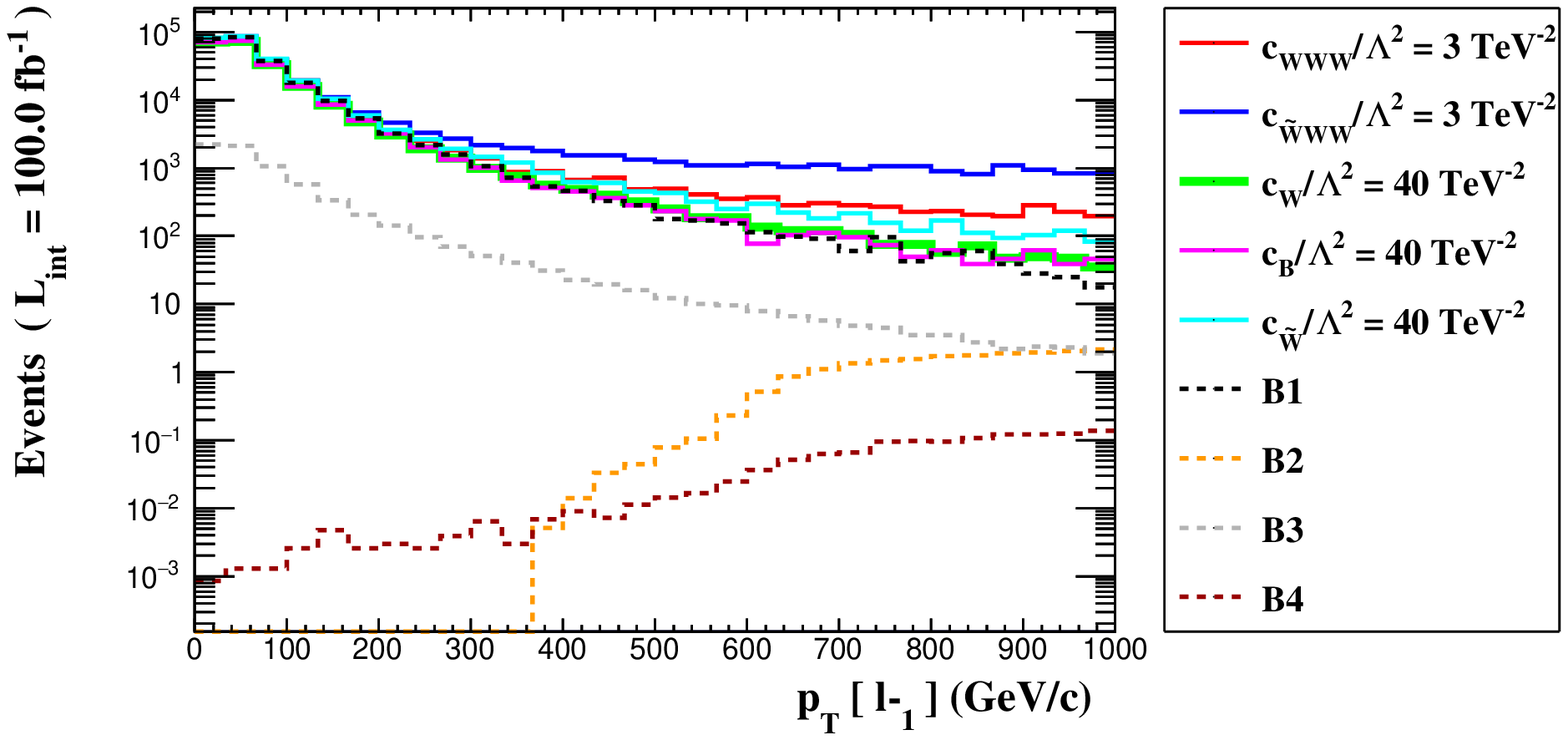}
\caption{}
\label{fig4:d}
\end{subfigure}\hfill

\begin{subfigure}{0.5\linewidth}
\includegraphics[width=\linewidth]{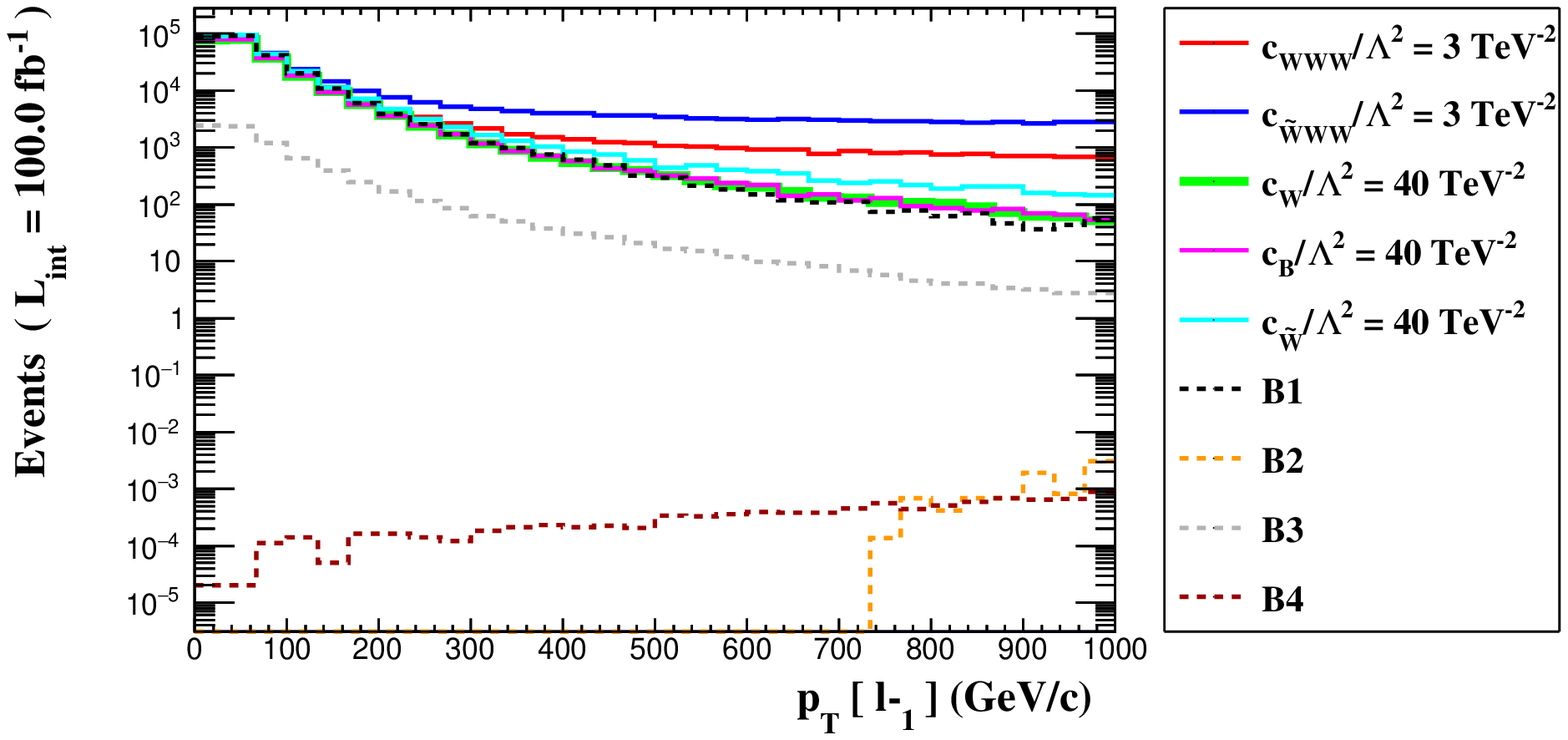}
\caption{}
\label{fig4:e}
\end{subfigure}\hfill
\caption{The number of events as a function of the leading charged lepton transverse momentum for the signal process and their relevant backgrounds with center-of-mass energies of (a) $\sqrt{s}=3$ TeV, (b) $\sqrt{s}=6$ TeV, (c) $\sqrt{s}=10$ TeV, (d) $\sqrt{s}=14$ TeV and (e) $\sqrt{s}=30$ TeV}
\label{fig4}
\end{figure}

In Fig.~\ref{fig4}, the number of events as a function of the $p_T^\ell$ leading charged lepton transverse momentum for the signal process and their relevant backgrounds have been computed at $\sqrt{s}=3, 6, 10, 14$ and $30$ TeV and ${\cal L}_{\text{int}}=100$ fb$^{-1}$ applying minimum cuts such as transverse momentum $p_T^{\gamma}>10$ GeV, pseudo-rapidity $|\eta^{\ell , \gamma}|<2.5$, minimum distance between leptons $\Delta R^{\ell\ell}_{\text{min}}>0.4$ and minimum distance between photons and leptons $\Delta R^{\gamma \ell}_{\text{min}}>0.4$. In this analysis, the $p_T^\ell$ leading charged lepton transverse momentums are used as a tool to probe sensitivity of the anomalous $c_{WWW}/{\Lambda^2}$, $c_{\widetilde{W}WW}/\Lambda^2$, $c_{B}/{\Lambda^2}$, $c_{W}/{\Lambda^2}$ and $c_{\widetilde{W}}/\Lambda^2$ couplings. As seen in the sub-figures in Fig.~\ref{fig4}, the deviation of the signal from the SM background ($B1$) for five anomalous couplings corresponds to $p_T^\ell=200$ GeV on average. As the center-of-mass energies of the muon collider increase, the $p_T^\ell$ transverse momentum value at which the deviation occurs decreases from about 300 GeV to 100 GeV. Therefore, in this study, we have performed the following final cut; $p_T^\ell >200$ GeV at Cut-3.

\section{Sensitivity analysis on the anomalous $c_{WWW}/{\Lambda^2}$, $c_{\widetilde{W}WW}/\Lambda^2$, $c_{B}/{\Lambda^2}$, $c_{W}/{\Lambda^2}$ and $c_{\widetilde{W}}/\Lambda^2$ couplings}

We have obtained 95$\%$ confidence level (C.L.) limits on the anomalous $c_{WWW}/{\Lambda^2}$, $c_{\widetilde{W}WW}/\Lambda^2$, $c_{B}/{\Lambda^2}$, $c_{W}/{\Lambda^2}$ and $c_{\widetilde{W}}/\Lambda^2$ couplings using a $\chi^2$ analysis with systematic errors to determine the sensitivities at the multi-TeV muon collider. The $\chi^2$ function is as follows:

\begin{eqnarray}
\label{eq.20} 
\chi^2=\left(\frac{\sigma_{B_{tot}}-\sigma_{NP}}{\sigma_{B_{tot}}\sqrt{\left(\delta_{st}\right)^2+\left(\delta_{sys}\right)^2}}\right)^2
\end{eqnarray}

{\raggedright where $\sigma_{B_{tot}}$ is only the total cross section of total SM backgrounds ($B_{tot}$) and $\sigma_{NP}$ is the cross section containing contributions from presence of both new physics beyond the SM and total SM backgrounds ($B_{tot}$), defined by $B_{tot}=B1+B2+B3+B4$. $\delta_{st}=\frac{1}{\sqrt {N_{B_{tot}}}}$ and $\delta_{sys}$ are the statistical error and the systematic error, respectively. The number of events in the total SM backgrounds is determined with $N_{B_{tot}}={\cal L}_{\text{int}} \times \sigma_{B_{tot}}$, where ${\cal L}_{\text{int}}$ is the integrated luminosity.}

The 95$\%$ C.L. limits calculated by Eq.~(\ref{eq.20}) on the anomalous $c_{WWW}/{\Lambda^2}$, $c_{\widetilde{W}WW}/\Lambda^2$, $c_{B}/{\Lambda^2}$, $c_{W}/{\Lambda^2}$ and $c_{\widetilde{W}}/\Lambda^2$ couplings without and with systematic errors 3$\%$, 5$\%$ are shown in Tables~\ref{tab5}-\ref{tab9}. The obtained limits for the main process $\mu^+ \mu^-\,\rightarrow\,\mu^+\gamma^* \mu^-\,\rightarrow\,\mu^+\ell^-\nu_\ell \bar{\nu}_\ell$ at the muon collider with different center-of-mass energies and luminosities are given in Table~\ref{tab1}. In addition, changes in the sensitivities according to the variation of cuts are also examined in Tables~\ref{tab5}-\ref{tab9}.

Many sources such as detector luminosity, trigger efficiencies, jet energy calibration, b-jet tagging efficiencies, lepton identification, background, initial and final state radiation (ISR/FSR), parton distribution functions (PDF) cause systematic uncertainties \cite{Khoriauli:2008xza}. Therefore, the inclusion of systematic uncertainties in statistical methods is necessary in studies. Taking into account previous work \cite{Koksal:2019ybm}, in this study, we consider the systematic uncertainties of $0\%$, $3\%$ and $5\%$ on the anomalous $c_{WWW}/{\Lambda^2}$, $c_{\widetilde{W}WW}/\Lambda^2$, $c_{B}/{\Lambda^2}$, $c_{W}/{\Lambda^2}$ and $c_{\widetilde{W}}/\Lambda^2$ couplings.

\begin{table} [H]
\caption{For systematic errors of $0\%$, $3\%$ and $5\%$, the 95\% C.L. limits on the anomalous $c_{WWW}/\Lambda^2$ coupling with center-of-mass energies and cuts.}
\label{tab5}
\begin{ruledtabular}
\begin{tabular}{ccccc}
\multicolumn{2}{c}{} & \multicolumn{3}{c}{$c_{WWW}/\Lambda^2$ (TeV$^{-2}$)} \\
\hline
$\sqrt{s}$ & Cuts & $\delta_{sys}=0\%$ & $\delta_{sys}=3\%$ & $\delta_{sys}=5\%$ \\ 
\hline \hline
\multirow{2}{*}{3 TeV} 
 & Cut-1 & [-0.740; 0.738] & [-8.800; 9.218] & [-11.245; 10.186]\\ 
 & Cut-3 & [-0.618; 0.614] & [-2.094; 2.091] & [-2.781; 2.779]\\ \hline
\multirow{2}{*}{6 TeV} 
 & Cut-1 & [-0.611; 0.624] & [-5.755; 5.673] & [-7.417; 7.274]\\ 
 & Cut-3 & [-0.304; 0.318] & [-1.290; 1.306] & [-1.666; 1.683]\\ \hline
\multirow{2}{*}{10 TeV}  
 & Cut-1 & [-0.357; 0.267] & [-3.630; 3.616] & [-4.592; 4.619]\\ 
 & Cut-3 & [-0.136; 0.156] & [-0.804; 0.824] & [-1.041; 1.060]\\ \hline
\multirow{2}{*}{14 TeV} 
 & Cut-1 & [-0.188; 0.170] & [-2.595; 2.576] & [-3.332; 3.311]\\ 
 & Cut-3 & [-0.080; 0.094] & [-0.586; 0.600] & [-0.759; 0.773]\\ \hline
\multirow{2}{*}{30 TeV} 
 & Cut-1 & [-0.058; 0.052] & [-1.201; 1.197] & [-1.550; 1.545]\\ 
 & Cut-3 & [-0.030; 0.025] & [-0.286; 0.281] & [-0.369; 0.364]\\ 
\end{tabular}
\end{ruledtabular}
\end{table}

\begin{table} [H]
\caption{Same as Table~\ref{tab5} but for the anomalous $c_{\widetilde{W}WW}/\Lambda^2$ coupling.}
\label{tab6}
\begin{ruledtabular}
\begin{tabular}{ccccc}
\multicolumn{2}{c}{} & \multicolumn{3}{c}{$c_{\widetilde{W}WW}/\Lambda^2$ (TeV$^{-2}$)} \\
\hline
$\sqrt{s}$ & Cuts & $\delta_{sys}=0\%$ & $\delta_{sys}=3\%$ & $\delta_{sys}=5\%$ \\ 
\hline \hline
\multirow{2}{*}{3 TeV} 
 & Cut-1 & [-0.818; 1.303] & [-5.386; 5.292] & [-6.535; 6.324]\\ 
 & Cut-3 & [-0.457; 0.470] & [-1.202; 1.215] & [-1.547; 1.561]\\ \hline
\multirow{2}{*}{6 TeV} 
 & Cut-1 & [-0.334; 0.289] & [-2.915; 2.890] & [-3.756; 3.747]\\ 
 & Cut-3 & [-0.150; 0.160] & [-0.644; 0.654] & [-0.833; 0.842]\\ \hline
\multirow{2}{*}{10 TeV}  
 & Cut-1 & [-0.163; 0.133] & [-1.782; 1.756] & [-2.293; 2.270]\\ 
 & Cut-3 & [-0.077; 0.070] & [-0.411; 0.404] & [-0.529; 0.522]\\ \hline
\multirow{2}{*}{14 TeV} 
 & Cut-1 & [-0.081; 0.095] & [-1.271; 1.284] & [-1.642; 1.655]\\ 
 & Cut-3 & [-0.046; 0.041] & [-0.299; 0.294] & [-0.385; 0.380]\\ \hline
\multirow{2}{*}{30 TeV} 
 & Cut-1 & [-0.030; 0.025] & [-0.600; 0.595] & [-0.774; 0.769]\\ 
 & Cut-3 & [-0.014; 0.014] & [-0.142; 0.142] & [-0.183; 0.183]\\ 
\end{tabular}
\end{ruledtabular}
\end{table}

\begin{table} [H]
\caption{Same as Table~\ref{tab5} but for the anomalous $c_{B}/\Lambda^2$ coupling.}
\label{tab7}
\begin{ruledtabular}
\begin{tabular}{ccccc}
\multicolumn{2}{c}{} & \multicolumn{3}{c}{$c_{B}/\Lambda^2$ (TeV$^{-2}$)} \\
\hline
$\sqrt{s}$ & Cuts & $\delta_{sys}=0\%$ & $\delta_{sys}=3\%$ & $\delta_{sys}=5\%$ \\ 
\hline \hline
\multirow{2}{*}{3 TeV} 
 & Cut-1 & [-4.089; 4.092] & [-19.107; 111.805] & [-31.514; 131.197]\\ 
 & Cut-3 & [-3.290; 3.736] & [-17.979; 92.478] & [-26.788; 101.596]\\ \hline
\multirow{2}{*}{6 TeV} 
 & Cut-1 & [-2.222; 2.222] & [-18.757; 83.925] & [-30.510; 91.708]\\ 
 & Cut-3 & [-1.261; 1.324] & [-17.011; 64.371] & [-25.165; 70.890]\\ \hline
\multirow{2}{*}{10 TeV} 
 & Cut-1 & [-1.134; 1.134] & [-18.582; 77.292] & [-30.090; 82.278]\\ 
 & Cut-3 & [-0.668; 0.691] & [-15.341; 55.858] & [-22.634; 63.262]\\ \hline
\multirow{2}{*}{14 TeV} 
 & Cut-1 & [-1.089; 1.089] & [-17.741; 71.719] & [-29.668; 76.532]\\ 
 & Cut-3 & [-0.437; 0.447] & [-14.898; 50.431] & [-21.914; 56.779]\\ \hline
\multirow{2}{*}{30 TeV} 
 & Cut-1 & [-1.040; 1.040] & [-17.163; 63.727] & [-29.354; 68.263]\\ 
 & Cut-3 & [-0.185; 0.187] & [-13.785; 41.720] & [-20.012; 47.842]\\
\end{tabular}
\end{ruledtabular}
\end{table}

\begin{table} [H]
\caption{Same as Table~\ref{tab5} but for the anomalous $c_{W}/\Lambda^2$ coupling.}
\label{tab8}
\begin{ruledtabular}
\begin{tabular}{ccccc}
\multicolumn{2}{c}{} & \multicolumn{3}{c}{$c_{W}/\Lambda^2$ (TeV$^{-2}$)} \\
\hline
$\sqrt{s}$ & Cuts & $\delta_{sys}=0\%$ & $\delta_{sys}=3\%$ & $\delta_{sys}=5\%$ \\ 
\hline \hline
\multirow{2}{*}{3 TeV} 
 & Cut-1 & [-3.997; 3.999] & [-19.261; 95.902] & [-31.823; 123.895]\\ 
 & Cut-3 & [-3.404; 3.793] & [-18.850; 76.918] & [-27.936; 83.943]\\ \hline
\multirow{2}{*}{6 TeV} 
 & Cut-1 & [-2.228; 2.228] & [-18.657; 83.402] & [-30.680; 91.947]\\ 
 & Cut-3 & [-1.268; 1.337] & [-16.849; 64.081] & [-24.884; 71.425]\\ \hline
\multirow{2}{*}{10 TeV} 
 & Cut-1 & [-1.134; 1.134] & [-18.450; 77.452] & [-29.884; 82.451]\\ 
 & Cut-3 & [-0.669; 0.690] & [-15.621; 55.227] & [-23.049; 61.351]\\ \hline
\multirow{2}{*}{14 TeV} 
 & Cut-1 & [-1.091; 1.091] & [-17.729; 71.604] & [-29.533; 76.046]\\ 
 & Cut-3 & [-0.432; 0.443] & [-14.601; 50.473] & [-21.466; 57.479]\\ \hline
\multirow{2}{*}{30 TeV} 
 & Cut-1 & [-1.040; 1.040] & [-17.258; 63.659] & [-29.363; 68.910]\\ 
 & Cut-3 & [-0.187; 0.189] & [-13.906; 41.779] & [-20.160; 47.769]\\ 
\end{tabular}
\end{ruledtabular}
\end{table}

\begin{table} [H]
\caption{Same as Table~\ref{tab5} but for the anomalous $c_{\widetilde{W}}/\Lambda^2$ coupling.}
\label{tab9}
\begin{ruledtabular}
\begin{tabular}{ccccc}
\multicolumn{2}{c}{} & \multicolumn{3}{c}{$c_{\widetilde{W}}/\Lambda^2$ (TeV$^{-2}$)} \\
\hline
$\sqrt{s}$ & Cuts & $\delta_{sys}=0\%$ & $\delta_{sys}=3\%$ & $\delta_{sys}=5\%$ \\ 
\hline \hline
\multirow{2}{*}{3 TeV} 
 & Cut-1 & [-8.593; 7.531] & [-47.762; 43.447] & [-66.136; 56.158]\\ 
 & Cut-3 & [-7.819; 7.288] & [-19.939; 19.394] & [-25.552; 24.995]\\ \hline
\multirow{2}{*}{6 TeV} 
 & Cut-1 & [-5.641; 4.705] & [-43.781; 42.261] & [-56.509; 55.298]\\ 
 & Cut-3 & [-3.792; 4.025] & [-16.228; 16.433] & [-20.979; 21.165]\\ \hline
\multirow{2}{*}{10 TeV} 
 & Cut-1 & [-5.071; 3.109] & [-42.242; 39.923] & [-55.799; 54.203]\\ 
 & Cut-3 & [-2.659; 2.564] & [-14.626; 14.532] & [-18.869; 18.776]\\ \hline
\multirow{2}{*}{14 TeV} 
 & Cut-1 & [-3.808; 2.946] & [-40.068; 38.745] & [-51.126; 50.139]\\ 
 & Cut-3 & [-1.948; 2.014] & [-13.550; 13.592] & [-17.513; 17.537]\\ \hline
\multirow{2}{*}{30 TeV} 
 & Cut-1 & [-3.228; 1.841] & [-37.018; 35.020] & [-47.366; 45.629]\\ 
 & Cut-3 & [-1.177; 1.097] & [-11.977; 11.887] & [-15.453; 15.358]\\
\end{tabular}
\end{ruledtabular}
\end{table}

In Tables~\ref{tab5}-\ref{tab9}, the sensitivities are investigated according to anomalous couplings, center-of-mass energies of the muon collider. If we first consider anomalous couplings with the same center-of-mass energies and cuts, the sensitivities of the anomalous $c_{\widetilde{W}WW}/\Lambda^2$ coupling are higher than those of other anomalous couplings. Second, increasing the center-of-mass energies of the muon collider for each anomalous coupling increases the sensitivities. Third, the sensitivities increase with increasing suppression from Cut-1 to Cut-3 in all analyses. According to Tables~\ref{tab5}-\ref{tab9}, the multi-TeV muon collider with $\sqrt{s}=30$ TeV and ${\cal L}_{\text{int}}=90$ ab$^{-1}$ at the Cut-3 has the most sensitive limits on the anomalous couplings. These limits are as follows;

\begin{eqnarray}
\label{eq.21} 
c_{WWW}/\Lambda^2=[-0.030; 0.025]\,\text{TeV}^{-2}\,,
\end{eqnarray}
\begin{eqnarray}
\label{eq.22} 
c_{\widetilde{W}WW}/\Lambda^2=[-0.014; 0.014]\,\text{TeV}^{-2}\,,
\end{eqnarray}
\begin{eqnarray}
\label{eq.23} 
c_{B}/\Lambda^2=[-0.185; 0.187]\,\text{TeV}^{-2}\,,
\end{eqnarray}
\begin{eqnarray}
\label{eq.24} 
c_{W}/\Lambda^2=[-0.187; 0.189]\,\text{TeV}^{-2}\,,
\end{eqnarray}
\begin{eqnarray}
\label{eq.25} 
c_{\widetilde{W}}/\Lambda^2=[-1.177; 1.097]\,\text{TeV}^{-2}\,.
\end{eqnarray}

\begin{figure} [H]
\centering
\begin{subfigure}{0.5\textwidth}
\includegraphics[scale=0.77]{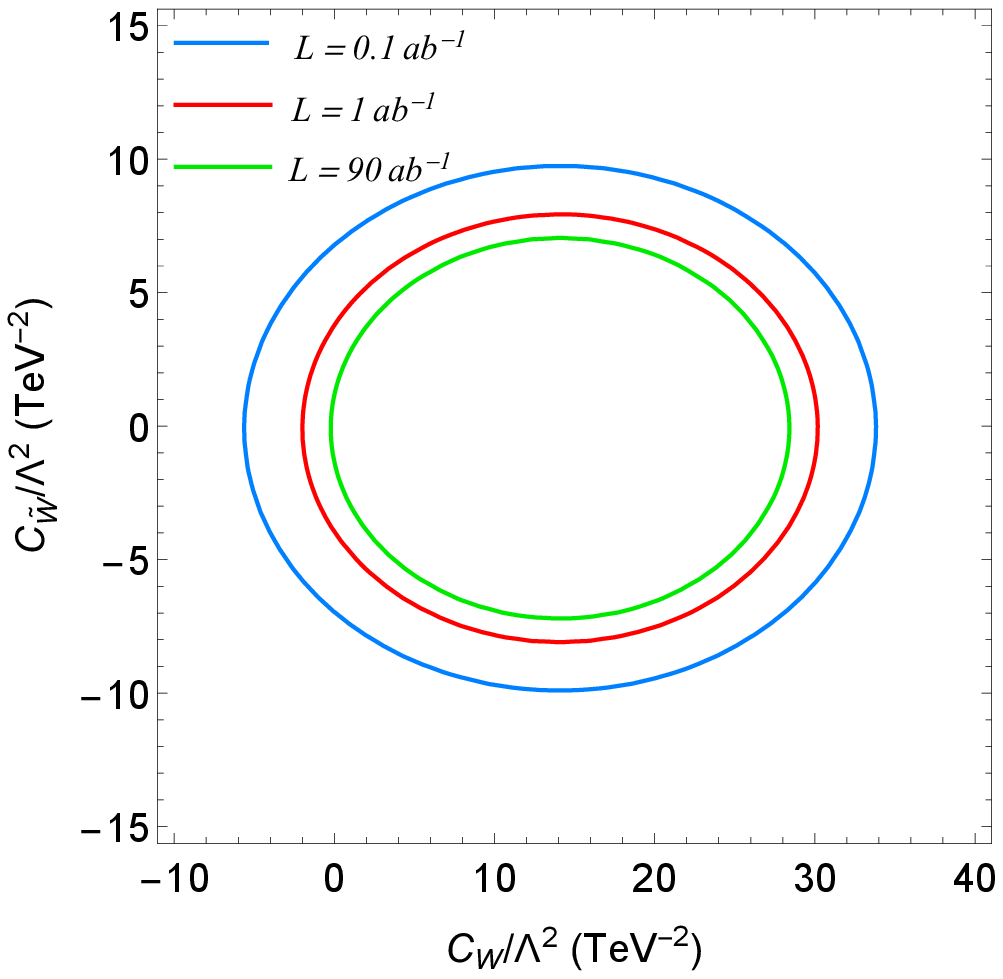}
\caption{}
\label{fig5:a}
\end{subfigure}\hfill
\begin{subfigure}{0.5\textwidth}
\includegraphics[scale=0.75]{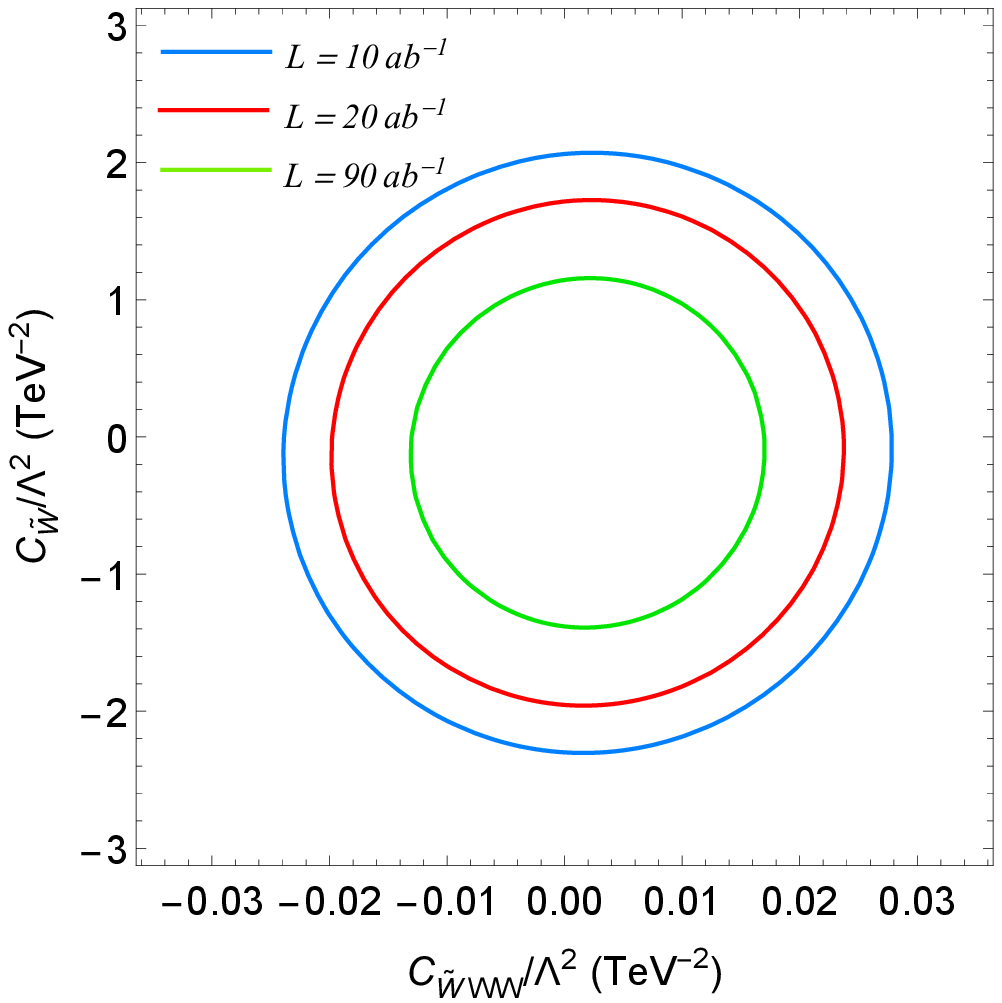}
\caption{}
\label{fig5:b}
\end{subfigure}\hfill

\begin{subfigure}{0.5\textwidth}
\includegraphics[scale=0.8]{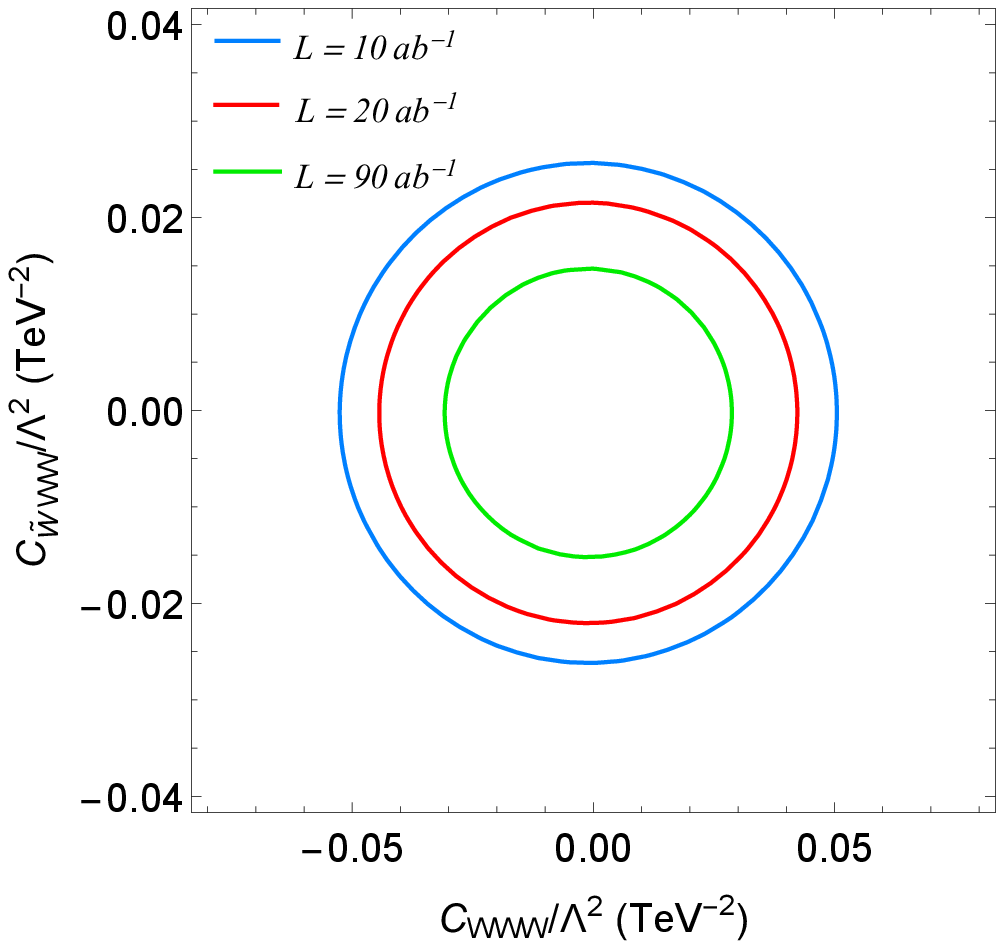}
\caption{}
\label{fig5:c}
\end{subfigure}\hfill
\caption{Two-dimensional sensitivity contours at the 95$\%$ C.L. in the (a) $c_{\widetilde{W}}/{\Lambda^2}$ - $c_{W}/{\Lambda^2}$ plane, (b) $c_{\widetilde{W}}/{\Lambda^2}$ - $c_{\widetilde{W}WW}/{\Lambda^2}$ plane and (c) $c_{\widetilde{W}WW}/{\Lambda^2}$ - $c_{WWW}/{\Lambda^2}$ plane according to different integrated luminosities at $\sqrt{s}=30$ TeV.}
\label{fig5}
\end{figure}

Finally, the contours for the anomalous couplings for the process $\mu^+ \mu^-\,\rightarrow\,\mu^+\gamma^* \mu^-\,\rightarrow\,\mu^+\ell^-\nu_\ell \bar{\nu}_\ell$ at the future muon collider for various integrated luminosities and a center-of-mass energy of 30 TeV which we obtained the best limit values only for the anomalous couplings are presented in Fig.~\ref{fig5}. In Figs. 5b and 5c, the integrated luminosities are 10, 20 and 90 ab$^{-1}$, whereas in Fig. 5a the different integrated luminosities from the other two figures are taken into account as 0.1, 1 and 90 ab$^{-1}$, because if the same integrated luminosities are considered, the two-dimensional contours would be very similar to each other. As we can see from these figures, the improvement in the sensitivity on the anomalous couplings is achieved by increasing to higher luminosities.

\section{Conclusions}

Our current study of the multi-TeV muon collider complements and even extends previous sensitivity estimates for the anomalous $c_{WWW}/{\Lambda^2}$, $c_{\widetilde{W}WW}/\Lambda^2$, $c_{B}/{\Lambda^2}$, $c_{W}/{\Lambda^2}$ and $c_{\widetilde{W}}/\Lambda^2$ couplings made in various specific colliders. It provides useful insights for planning the construction of future muon colliders. As a result, a muon collider in the high energy region with high integrated luminosity can both explore the wider parameter space and achieve better sensitivity couplings.

Experimental limits of the anomalous $WW\gamma$ coupling at the CMS Collaboration at the CERN LHC \cite{Sirunyan:2021rwx,Sirunyan:2019umc} are the most current reference point to compare our results. When our most sensitive limits given in Eqs.~(\ref{eq.21})-(\ref{eq.25}) are compared with the experimental limits in Refs.~\cite{Sirunyan:2021rwx,Sirunyan:2019umc}, it is seen that our sensitives on the anomalous couplings are between 12-47 times of magnitude better than the current experimental limits.

The multi-TeV muon colliders, which have the advantages of cleaner physics conditions in the detector and lower cost, create an attractive motivation with remarkable sensitivity limits. The success of the multi-TeV muon colliders suggests that such studies are definitely worth doing.

\section*{Declarations}
A preprint of this article was originally posted on the arXiv \cite{Spor:2022tef}. 
\begin{itemize}
\item Competing interests: The authors declare there are no competing interests.
\item Funding: The authors declare no specific funding for this work.
\item Data Availability: This manuscript does not report data.
\end{itemize}

\end{document}